\documentclass{article}
\usepackage[a4paper, total={6.5in,9in}]{geometry}
\usepackage[utf8]{inputenc}
\usepackage{amsmath}
\usepackage{amssymb}
\usepackage{graphicx}
\usepackage{authblk}
\usepackage{url}
\usepackage[authoryear]{natbib}
\usepackage{adjustbox}
\usepackage{rotating}
\usepackage{comment}
\usepackage{longtable}
\usepackage{xcolor}
\usepackage{ulem}
\usepackage[english]{babel}
\usepackage{setspace}
\usepackage{multirow}
\usepackage{amsfonts}
\usepackage{array} 
\usepackage{flushend} 
\usepackage{stfloats} 
\usepackage{color}
\usepackage{chngpage} 
\usepackage{totcount} 
\usepackage{hyperref}
\usepackage{fix-cm} 
\usepackage{algorithm}
\usepackage{algorithmicx} 
\usepackage{algpseudocode} 
\usepackage{listings} 
\usepackage{crop}
\usepackage{subfloat} 
\usepackage{subfig}
\usepackage{tikz} 
\usepackage{hyperref} 
\usepackage{footnote}
\usepackage{mathrsfs} 
\usepackage{wrapfig} 
\usepackage{amsthm} \usepackage[most]{tcolorbox}
\usepackage{listings}  
\tcbuselibrary{listings, breakable}

\newtcolorbox[auto counter, number within=section]{mybox}[2][]{%
	enhanced,
	colback=blue!5!white,
	colframe=blue!75!black,
	fonttitle=\bfseries,
	coltitle=white,
	title=Box~\thetcbcounter: #2,
	listing only,
	listing options={language=R},
	unbreakable,
	boxsep=1pt,
	left=2pt,      
	right=2pt,
	top=2pt,
	bottom=2pt,
	sharp corners,
	#1
}

\begin{document}
	
	\vspace{-40pt}
	\title{An Introduction to Topological Data Analysis Ball Mapper in Python}
	\vspace{-30pt}

	\author[1]{Simon Rudkin \thanks{Full Address: Department of Social Statistics, School of Social Sciences, University of Manchester, Oxford Road, Manchester, M13 9PL, United Kingdom. Email:simon.rudkin@manchester.ac.uk}}
	\affil[1]{School of Social Sciences, University of Manchester, United Kingdom}
	
	\vspace{-40pt}
	\vspace{-20pt}
	\date{}
	\doublespacing
	
	\maketitle
	\vspace{-30pt}
	\begin{abstract}		
		Visualization of data is an important step in the understanding of data and the evaluation of statistical models. Topological Data Analysis Ball Mapper (TDABM) after \cite{dlotko2019ball}, provides a model free means to visualize multivariate datasets without information loss. To permit the construction of a TDABM graph, each variable must be ordinal and have sufficiently many values to make a scatterplot of interest. Where a scatterplot works with two, or three, axes, the TDABM graph can handle any number of axes simultaneously. The result is a visualization of the structure of data. The TDABM graph also permits coloration by additional variables, enabling the mapping of outcomes across the joint distribution of axes. The strengths of TDABM for understanding data, and evaluating models, lie behind a rapidly expanding literature. This guide provides an introduction to TDABM with code in Python. 
	\end{abstract}


\section{Introduction}

Consider data as a point cloud. The cloud has $K$ dimensions, with each data point $x_i \in X$ having co-ordinates $x_{ik}$ on each $k \in K$. The location of point $i$, $i \in [1,N]$ in the point cloud is analogous to the placement of point $i$ onto a scatter plot. The notion of a point cloud allows us to consider the appearance of data in cases where $K>2$\footnote{The case where $K=3$ can be visualized with a three dimensional scatterplot, but the reading of three-dimensional scatter plots is not always straightforward. For cases where $K>3$, human abilities to interpret graphical representations create difficulties in creating a full illustration of the full dimensionality of the data.}. This guide presents a means to visualize the resulting point cloud as an abstract two-dimensional representation. We present the Topological Data Analysis Ball Mapper (TDABM) algorithm of \cite{dlotko2019ball}. The purpose of this guide is to provide an introduction to TDABM and to discuss the code required to implement TDABM on a user dataset. Throughout the provided code is as implemented in Python using the pyBallMapper library\footnote{The Python library pyBallMapper is available from \href{https://github.com/dioscuri-tda/pyBallMapper/tree/main}{https://github.com/dioscuri-tda/pyBallMapper/tree/main}. The code to recreate this guide is available as a Jupyter Notebook at \href{https://github.com/srudkin12/BM-Guide}{https://github.com/srudkin12/BM-Guide}.}. 
Consider a dataset $X$, which comprises the $K$ variables that are of interest. The first objective is to understand the relationships between the variables within $X$. \cite{matejka2017same} reminds on the importance of visualization of datasets. In addition to demonstrating correlation structures, visualization at the exploratory data analysis stage enables the identification of outliers, potential leverage points and other irregularities which may warrant further investigation. To the dataset, an outcome of interest, $Y$, is introduced. Mapper algorithms, like TDABM, aim to map $Y$ across the multidimensional space of $X$. In the two-dimensional scatterplot setting, $Y$ can be shown by coloring points. The mapping on a 2-dimensional case is attainable on a scatterplot by coloring the data points. As the challenge of simultaneously viewing the $X$ variables becomes harder as $K$ increases, so the ability to show the mapping of$Y$ also intensifies. It is the inability to map outcomes across multiple dimensions to which TDABM speaks.

Within the growing literature applying TDABM within Economics and Finance, data may be grouped into representations of time series and of cross sections. \cite{rudkin2024return} considers the time series of Bitcoin returns, taking each weekly return as the axis in the point cloud. With the contemporary value and 4 lags, the resulting point cloud has 5 dimension. Further considerations of time series include \cite{rudkin2023regional} and \cite{rudkin2023economic}. In the majority of published applications in Economics and Finance, take a dataset constructed of multiple variables, rather than a single time series. \cite{rudkin2023economic} take the socio-demographic characteristics of UK parliamentary constituencies as the axes of the cloud. \cite{tubadji2025cultural} and \cite{rudkin2024topology} similarly look at the joint distribution of regional socio-demographic characteristics, focusing on the European Union NUTS regions. In Finance the joint distribution of firm characteristics represent the inputs for \cite{dlotko2021financial} and \cite{qiu2020refining}. In all cases the first objective is to identify structure within the data. 

The diversity of outcomes of interest studied in the emerging literature is also notable. Papers such as \cite{rudkin2024return}, \cite{dlotko2021financial} and \cite{qiu2020refining} have a forecasting approach. A forecasting approach takes data from a current period and asks what will happen to $Y$ in the subsequent periods. For \cite{rudkin2024return}, the object is to predict the direction of Bitcoin returns in subsequent periods. \cite{rudkin2023regional} looks at how the trajectory of development predicts the subsequent resilience of the region to the Global Financial Crisis. The ability of TDABM to found a forecasting model is discussed in \cite{rudkin2024return} and \cite{charmpi2023topological}. Discussion of forecasting here is placed within the context of TDA-based representations of the point cloud. Our discussion is independent of the ability of time series of topological signals to improve forecast performance. See \cite{shultz2023applications} for more on wider applications of TDA.

The original mapper algorithm was proposed by \cite{singh2007topological}. The \cite{singh2007topological} mapper first creates a segmentation of the data using a mechanism such as cluster analysis. A lens function reduces the dimensionality of the data. Once reduced in dimension, the data is split into equally sized bins with overlap. The overlap between bins is used to construct edges between the discs that represent each bin. Determination of lens function, clustering algorithm, number of bins and bin overlap, mean that the original mapper algorithm has many choice parameters. Contrasting with the single choice parameter in TDABM, one of the advantages of TDABM is clear. Work on the stability of the mapper algorithm also confirms TDABM to have an advantage. There are other methods of visualizing multivariate data, such as T-SNE or Principal Components Analysis (PCA), but both require dimension reduction. Dimension reduction causes data loss and hence is not seen as preferable when a method exists that does not require dimension reduction. TDABM requires no reduction of dimensions and therefore can be considered superior to T-SNE or PCA.

This guide is presented over 4 sections. Section \ref{sec:data} delivers an exposition of the two artificial datasets used within the guide. An explanation of the implementation of TDABM on the example dataset is provided by Section \ref{sec:method}. Section \ref{sec:art} is the primary reference point for applying TDABM. Within Section \ref{sec:art}, I walk through the construction of the TDABM graph with the code echoing the implementation in Python. The Python implementation of TDABM is discussed in Section \ref{sec:pybm}. Finally, Section \ref{sec:summary} summarises the guide and provides suggestions for further work.

\section{Data}
\label{sec:data}

I use two datasets to illustrate TDABM within this guide. The code for the production of the dataset is provided within the accompanying .ipynb file. both of the datasets are bivariate, with $X_1 \sim U[0,1]$ and $X_2 \sim U[0,1]$. Both datasets have 1000 observations, $N=1000$. In the first dataset the correlation between $X_1$ and $X_2$ is 0, in the second dataset we have $Cor(X_1,X_2) = 0.5$. Like any distance based methodology, it is important to ensure that the values of each variable are on comparable scales. One means of ensuring that variables are on the same scale is to standardize. Both $X_1$ and $X_2$ are standardized.  

The code which is used to create the datasets is given in Box \ref{box:pycode}. The code can be readily changed to produce alternative correlations between the two variables. The \texttt{random.seed(101)} is selected in order to ensure that the correlation is close to 0.5 for the specific random values generated. Having a fixed seed ensures that the results are all replicable. The resulting summary statistics of $X_1$ and $X_2$ are provided in Table \ref{tab:data1}.

\begin{mybox}[label=box:pycode]{R Code for Dataset}
	An initial bivariate dataset is produced, beginning with the generation of the initial values of $X_1$ and $X_2$ from $U \left[0,1\right]$:
	\begin{lstlisting}[language=Python]
		np.random.seed(101)
		n = 1000
		X1 = np.random.uniform(size=n)
		X2 = np.random.uniform(size=n)
	\end{lstlisting}
	Because a given correlation is enabled for Dataset 2, the data is scaled
	\begin{lstlisting}[language=Python]
		X1_scaled = (X1 - np.mean(X1)) / np.std(X1)
		X2_scaled = (X2 - np.mean(X2)) / np.std(X2)
	\end{lstlisting}
	The data is bound into a \texttt{pandas} \texttt{DataFrame} object
	\begin{lstlisting}[language=Python]
		df1=pd.DataFrame({
			'X1': X1_scaled,
			'X2': X2_scaled
		})
	\end{lstlisting}
	For Dataset 2 a new variable is created with correlation targetting 0.5
	\begin{lstlisting}
		X2_adjusted = 0.5 * X1_scaled + np.sqrt(1 - 0.5**2) * X2_scaled
		df2 = pd.DataFrame({
			'X1': X1_scaled,
			'X2': X2_adjusted
		})
	\end{lstlisting}
\end{mybox}

\begin{table}
	\begin{center}
		\caption{Summary Statistics}
		\label{tab:data1}
		\begin{tabular}{l c c c c c c}
			\hline
			Statistic & \multicolumn{3}{l}{Dataset 1} & \multicolumn{3}{l}{Dataset 2}\\
			& $X_1$ & $X_2$ & $Y$ & $X_1$ & $X_2$& $Y$\\
			\hline
			X1	X2	Y
			
			Mean &	0.000 & 0.000 & 0.000 &	0.000 & 0.000 & 0.000 \\
			Standard deviation & 1.000 & 1.000 &	1.418 & 1.000 & 0.999 & 1.002 \\
			Minimum &	-1.738 & -1.749 &	-3.244 & -1.738 & -2.335 & -2.240 \\
			25\%&	-0.833 &	-0.852 & -0.988 & -0.833 & -0.774 & -0.761 \\
			50\% &	-0.028&	0.032 &	0.019& -0.028 & 0.051 & 0.009 \\
			75\% &	0.844&	0.859&	1.027 & 0.844 & 0.723 & 0.713 \\
			Maximum &	1.753 &	1.734 & 	3.400 & 1.753 & 2.308 & 2.339\\
			\hline
		\end{tabular}
	\end{center}
	\raggedright
	\footnotesize{Notes: Summary statistics for the two artificial bivariate datasets used in this paper. $Y$ is computed such that $Y = X_1 - X_2$. $X_1$ and $X_2$ are drawn independently from $U[0,1]$. Variables are standardized. In Dataset 2, $X_2$ is further adjusted to ensure correlation of 0.5 with $X_1$. $N=1000$. Replication code is included within the accompanying .ipynb file.}
\end{table}

\begin{figure}
	\begin{center}
		\caption{Datasets used in Examples}
		\label{fig:data}
		\begin{tabular}{c c}
			\includegraphics[width=7cm]{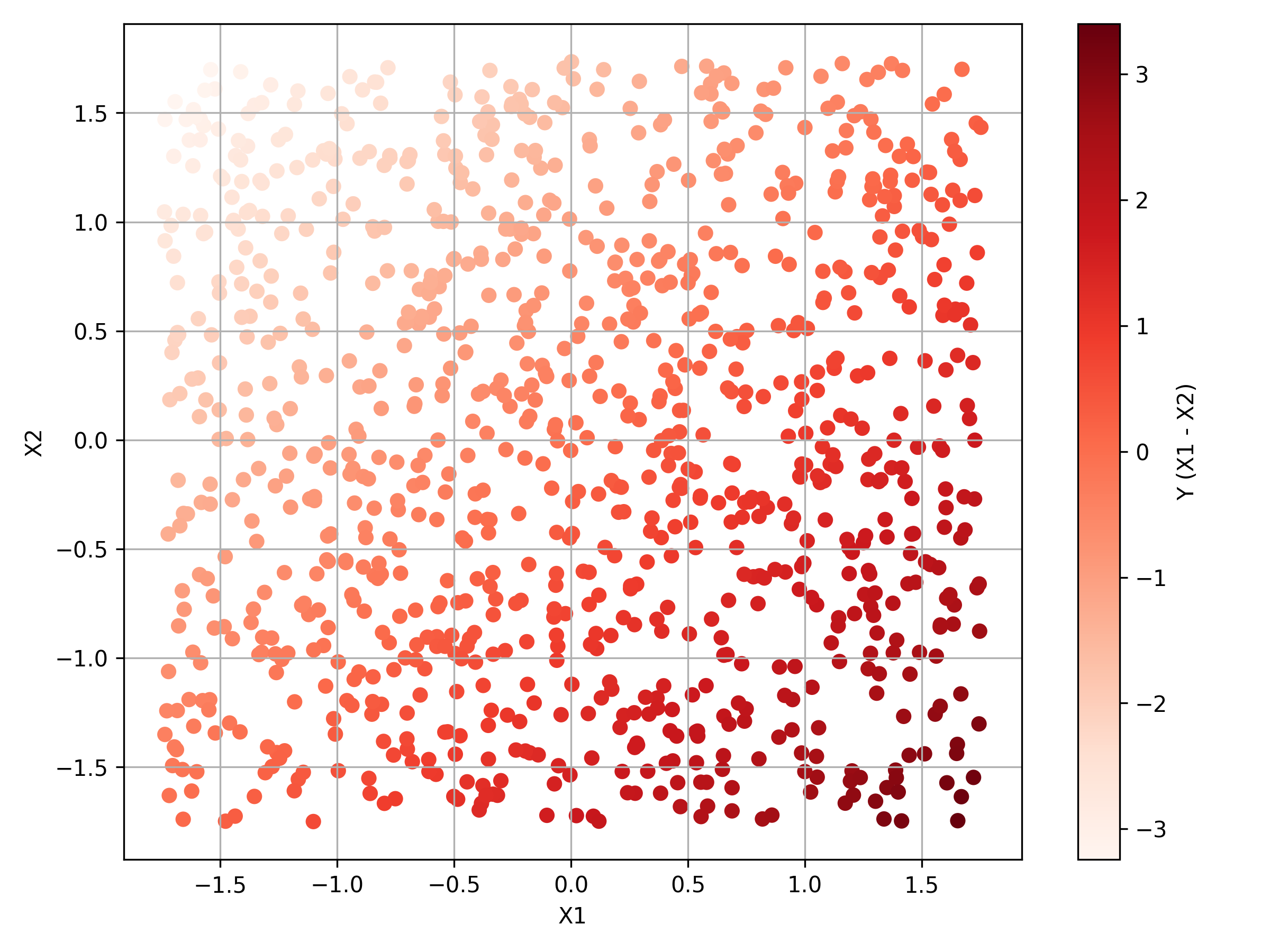}&
			\includegraphics[width=7cm]{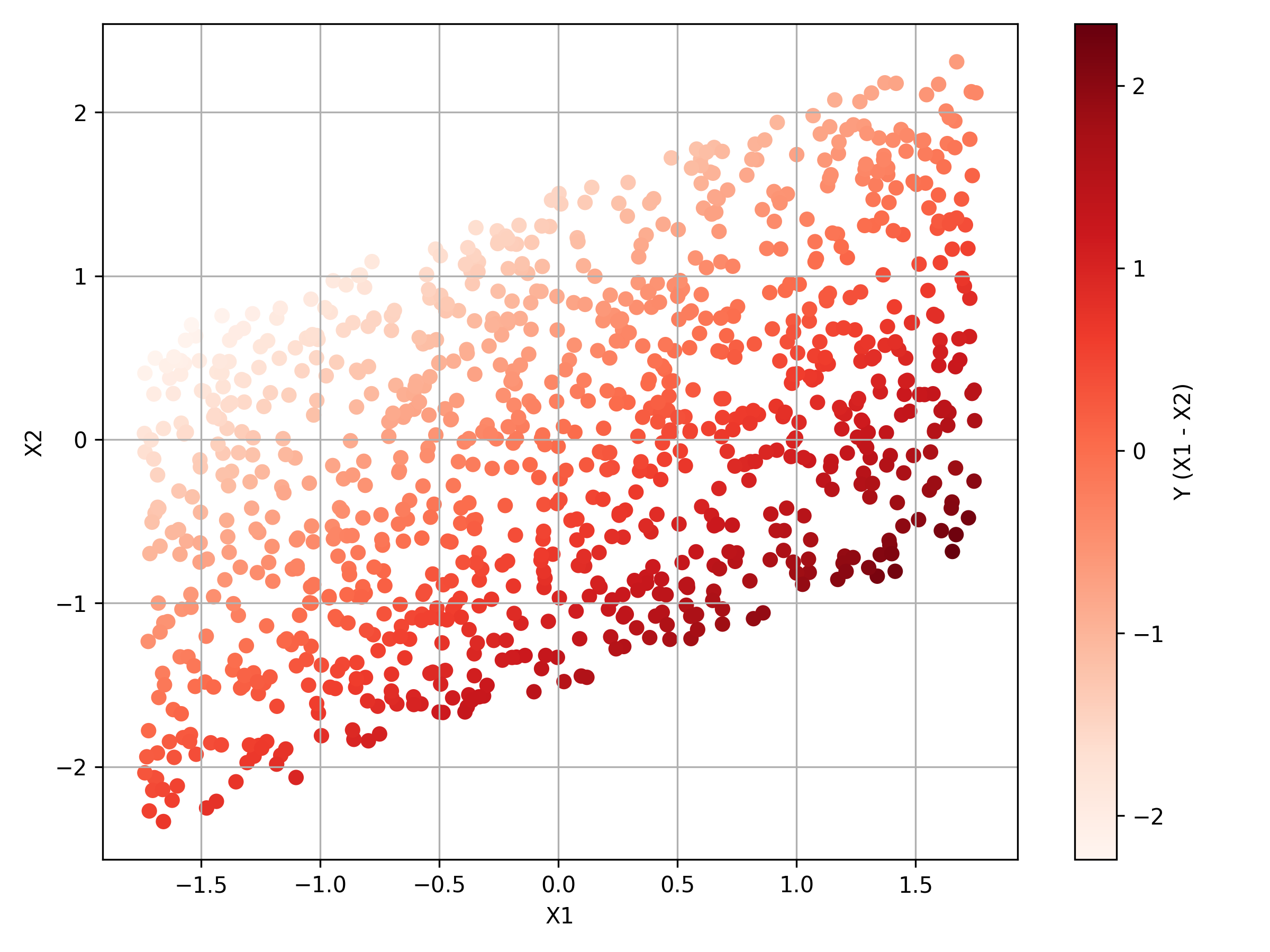}\\
			(a) Dataset 1 & (b) Dataset 2 \\
		\end{tabular}
	\end{center}
	\raggedright
		\footnotesize{Notes: Scatterplots for the two artificial bivariate datasets used in this paper. $Y$ is computed such that $Y = X_1 - X_2$. $X_1$ and $X_2$ are drawn independently from $U[0,1]$. Variables are standardized. In Dataset 2, $X_2$ is further adjusted to ensure correlation of 0.5 with $X_1$. $N=1000$. Coloration is according to the value of $Y$.}
\end{figure}

Table \ref{tab:data1} confirms that the summary statistics are in line with expectation. As the data has been standardised, the mean is 0 and standard deviation 1 for both $X_1$ and $X_2$. There are no discernable differences between the summary statistics for the two datasets, despite there being a correlation in Dataset 2. Both datasets share the same $X_1$ and so the summary statistics for $X_1$ in the first and fourth columns are identical. The correlation is confirmed as 0.496 in the output. To see the data, Figure \ref{fig:data} provides scatter plots for the two datasets.

The largest values of $Y$ appear when $X_1$ takes its' highest values and when $X_2$ is at its lowest values. Meanwhile the smallest values of $Y$ appear when $X_1$ is low and $X_2$ is high. The result is a diagonal pattern with the lowest values being to the upper left. Panel (a) shows that Dataset 1 has a clear pattern within the output. In panel (b), $X_1$ and $X_2$ are correlated and so the points are also arranged around the leading diagonal. The differentiation in the colours is harder to identify. However, because Python reassigns the colors for the points according to the specific dataset being plotted, the graduation between the lowest and highest values can still be seen within panel (b).

\section{Methodology}
\label{sec:method}

Within this guide there is a dataset $X$ with $K$ variables. For each point, $i$, $i \in \lbrace 1, ..., N \rbrace$, the location in the point cloud $P$ is set as $x_{ik}$, $k \in \lbrace 1, ..., K \rbrace$. In the simplest case a point cloud is similar to a scatter plot. For the examples used in this guide $K=2$, ensuring that the scatterplot is indeed analogous to the point cloud. In addition to $X$, the TDABM algorithm requires two further inputs. Firstly, a coloration variable which is available for all data points. Secondly, a radius for the balls in the cover, $\epsilon$. The radius performs a similar function to the scale when plotting a cartographic map. A small radius means that many details appear within the structure of the data. At a higher radius the global picture is apparent, but the microstructure of the data is obscured. In considering a dataset, trying multiple values of $\epsilon$ is recommended. Only after seeing the true data structure can a radius be chosen. TDABM creates a cover of $X$ allowing the abstract visualization of $X$ in two dimensions.

As written in \cite{dlotko2019ball}, the TDABM algorithm begins by choosing a point at random from $X$. The selected point becomes $l_1$, the first landmark. In Python the selection of the first landmark is coded as per Box \ref{box:tp1}. In the case of the first selection there is no need to consider whether any point is covered. Around the landmark a ball of radius $\epsilon$ is constructed. All points within the ball are considered to be covered by the ball that has been drawn around $l_1$. The ball around $l_1$ is labelled as Ball 1 and can be written as $B_1(X,\epsilon)$.  In the example $\epsilon=1.5$. The full code for the construction of Ball 1 is placed in Box \ref{box:tp1}.

\begin{mybox}[label=box:tp1]{First Ball Construction}
	Before beginning a random seed is set for reproducibility 
	\begin{lstlisting}[language=Python]
		np.random.seed(123)		
	\end{lstlisting}
	Selection of the random point that will be the first landmark is made using \texttt{numpy}
	\begin{lstlisting}[language=Python]
		random_index101 = np.random.choice(df1.index)
		random_point101 = df1.loc[random_index101]
	\end{lstlisting}
	Identify the distance of each data point from the first randomly selected point
	\begin{lstlisting}[language=Python]
		df1['distance'] = np.sqrt((df1['X1'] - random_point101['X1'])**2 + (df1['X2'] - random_point101['X2'])**2)
	\end{lstlisting}
	Find the points in \texttt{df1} with distance less than the ball radius. In this example the radius is 1.5
	\begin{lstlisting}[language=Python]
		radius = 1.5
		df1['inside_circle01'] = (df1['distance'] <= radius).astype(int)
	\end{lstlisting}
	Finally, a membership variable is constructed. This will take the value of the ball number if the point is only in one ball. Will progressively take the value 7 when the point is in two balls and 8 if the point is not in any balls. The example has 6 balls, hence the use of 7 and 8 as the next two integers.
	\begin{lstlisting}[language=Python]
		df1['circle_membership'] = np.where(df1['inside_circle01'] == 1, 1, 8)
	\end{lstlisting}
\end{mybox}

\begin{figure}
	\begin{center}
		\caption{Dataset 1 Random Cover}
		\label{fig:data1t}
		\begin{tabular}{c c}
			\includegraphics[width=7cm]{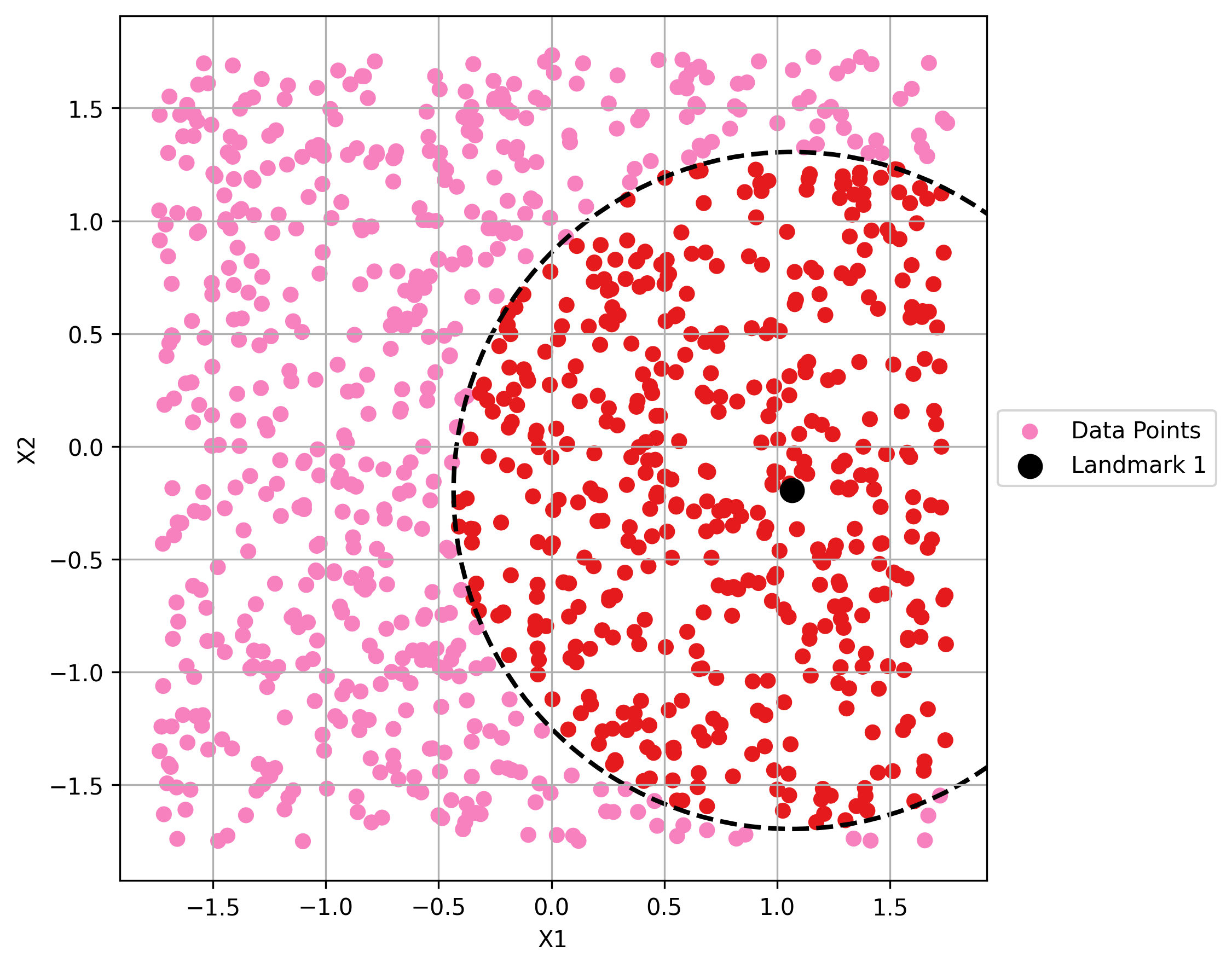}&
			\includegraphics[width=7cm]{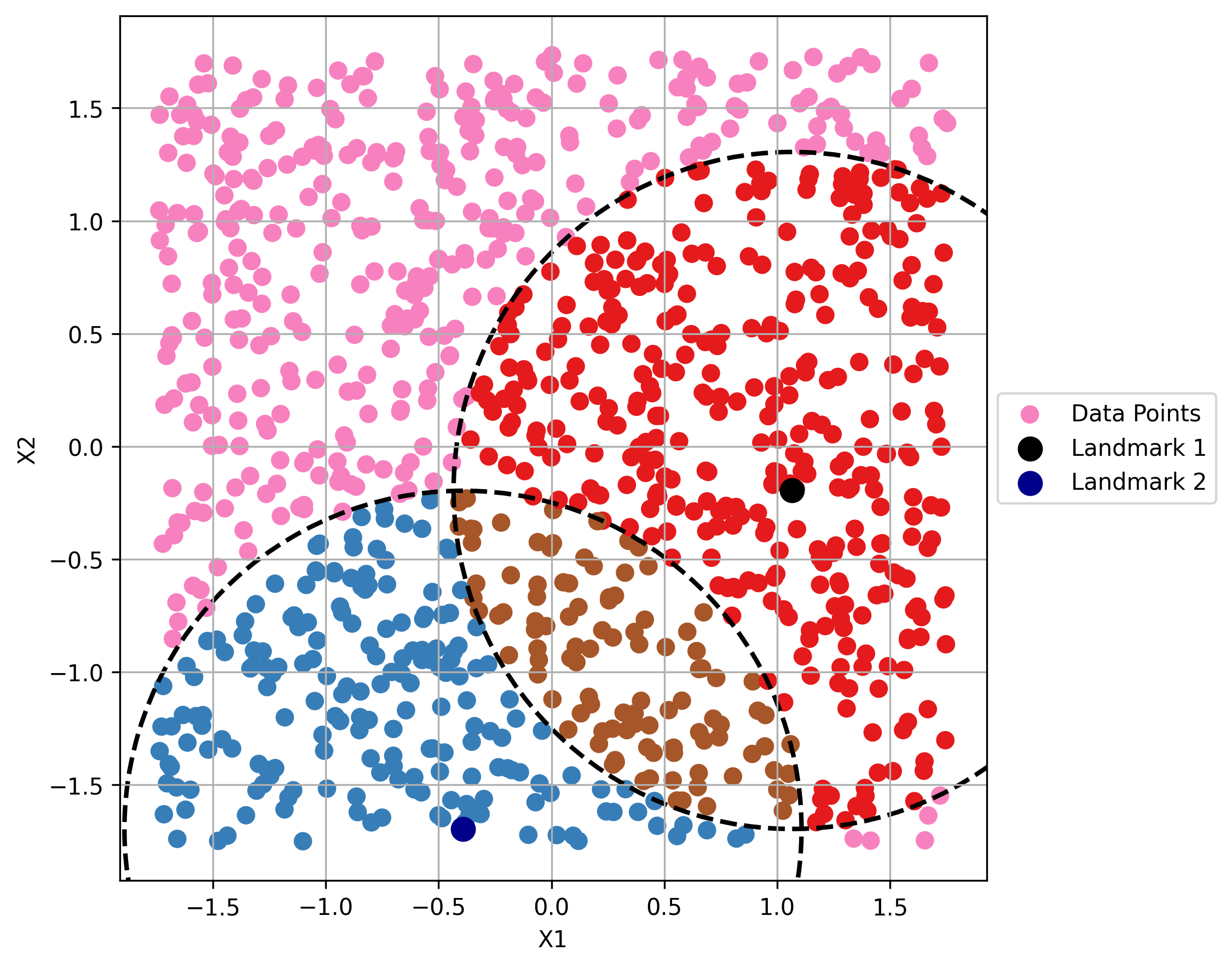}\\
			(a) Landmark 1 & (b) Landmark 2 \\
			\includegraphics[width=7cm]{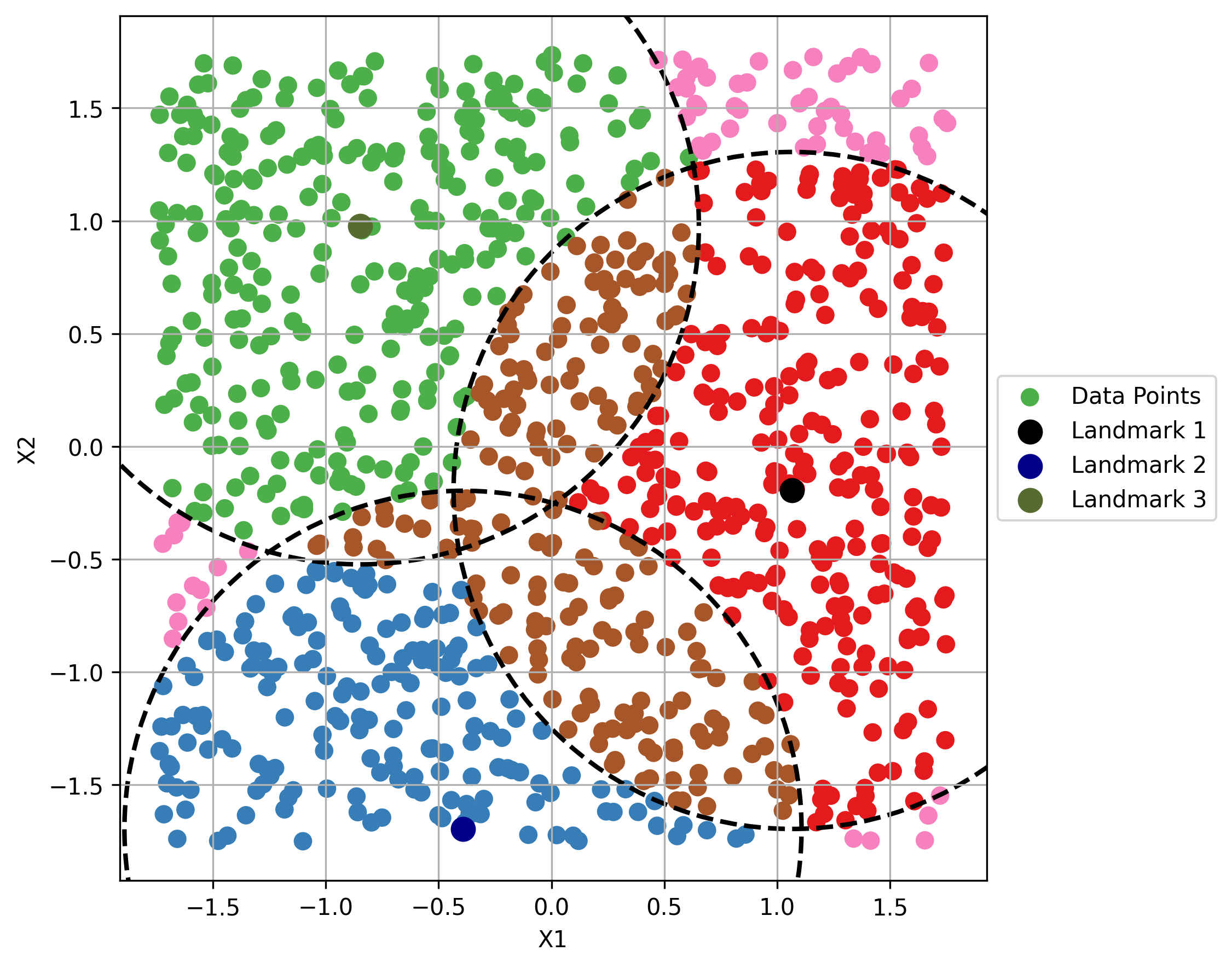}&
			\includegraphics[width=7cm]{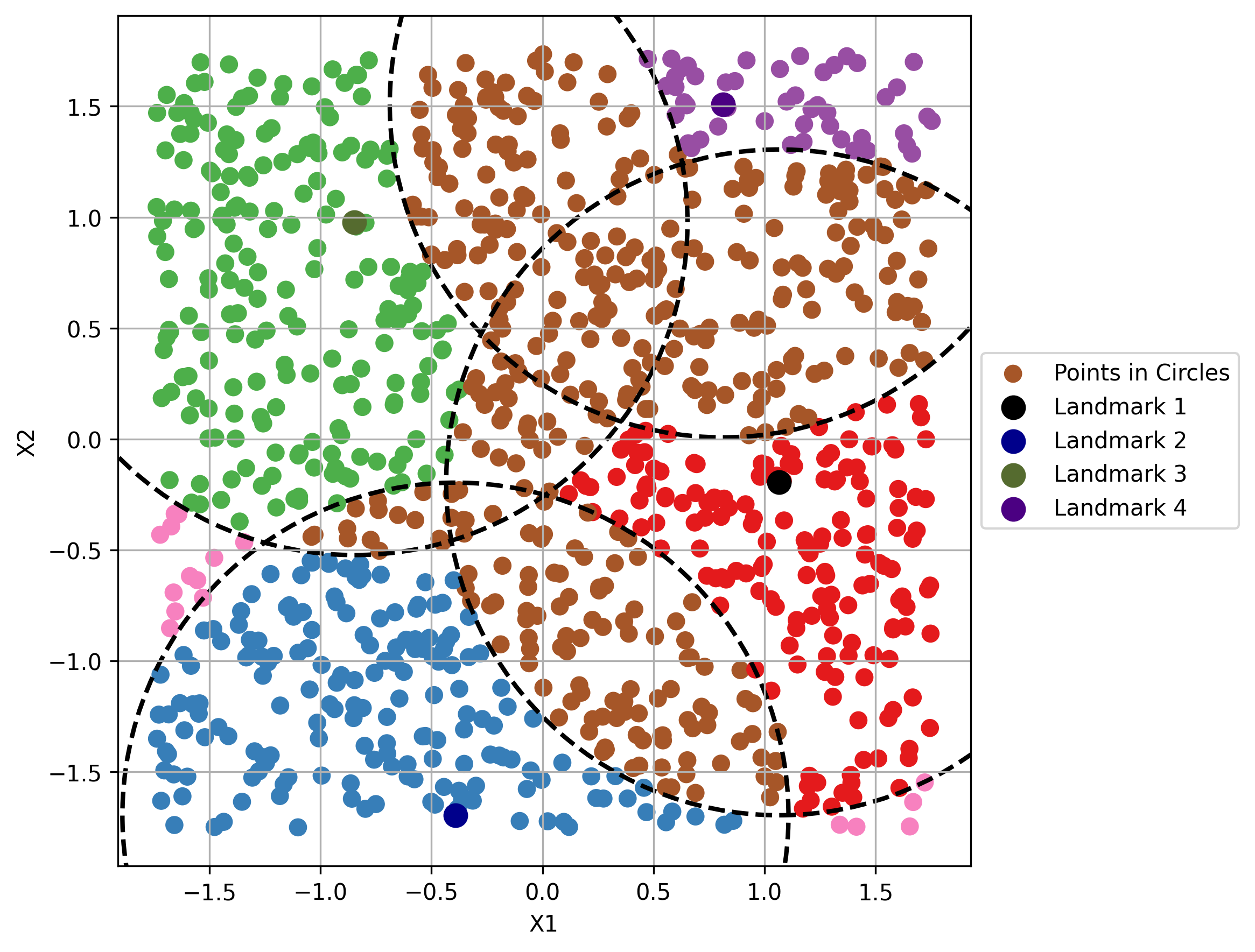}\\
			(c) Landmark 3 & (d) Landmark 4 \\
			\includegraphics[width=7cm]{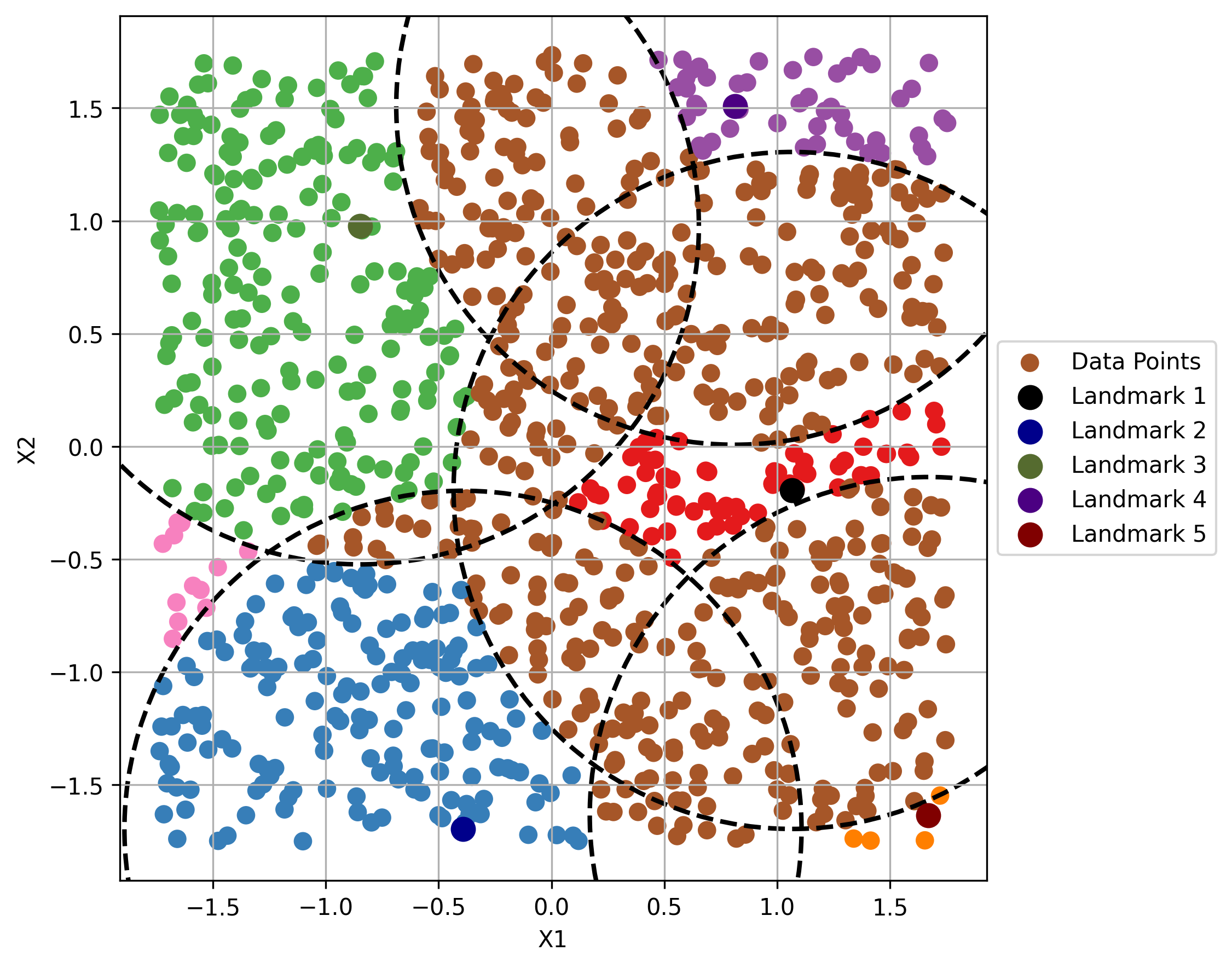}&
			\includegraphics[width=7cm]{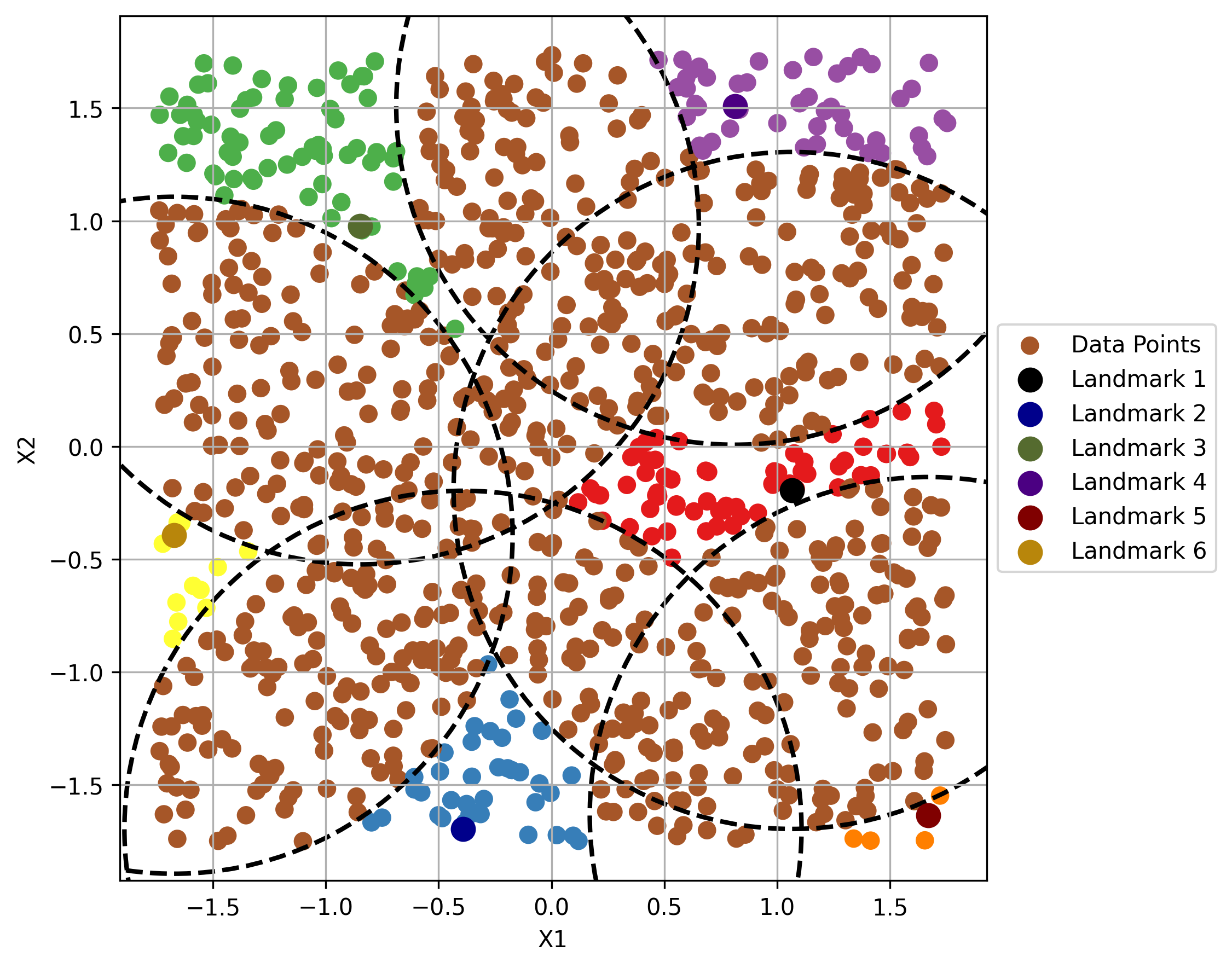}\\
			(e) Landmark 5 & (f) Landmark 6\\
			
		\end{tabular}
	\end{center}
	\raggedright
	\footnotesize{Notes: Figures represent the stepwise construction of the TDABM style plot. Landmarks refer to the points that would be the landmarks of the balls in a TDABM plot. Numbering of landmarks is the order of selection. Landmarks are selected from the uncovered points (pink) until all points can be found within at least one circle (all other colors). Dashed circles represent the boundaries of the balls around each landmark. Dataset 1 has two variables $X_1$ and $X_2$, and contains 1000 points. $X_1$ and $X_2$ are drawn at random from $U \left[0,1\right]$. Standardization is applied to allow comparison with Dataset 2.}
\end{figure}

\begin{figure}
	\begin{center}
		\caption{Dataset 2 Random Cover}
		\label{fig:data2t}
		\begin{tabular}{c c c}
			\includegraphics[width=5cm]{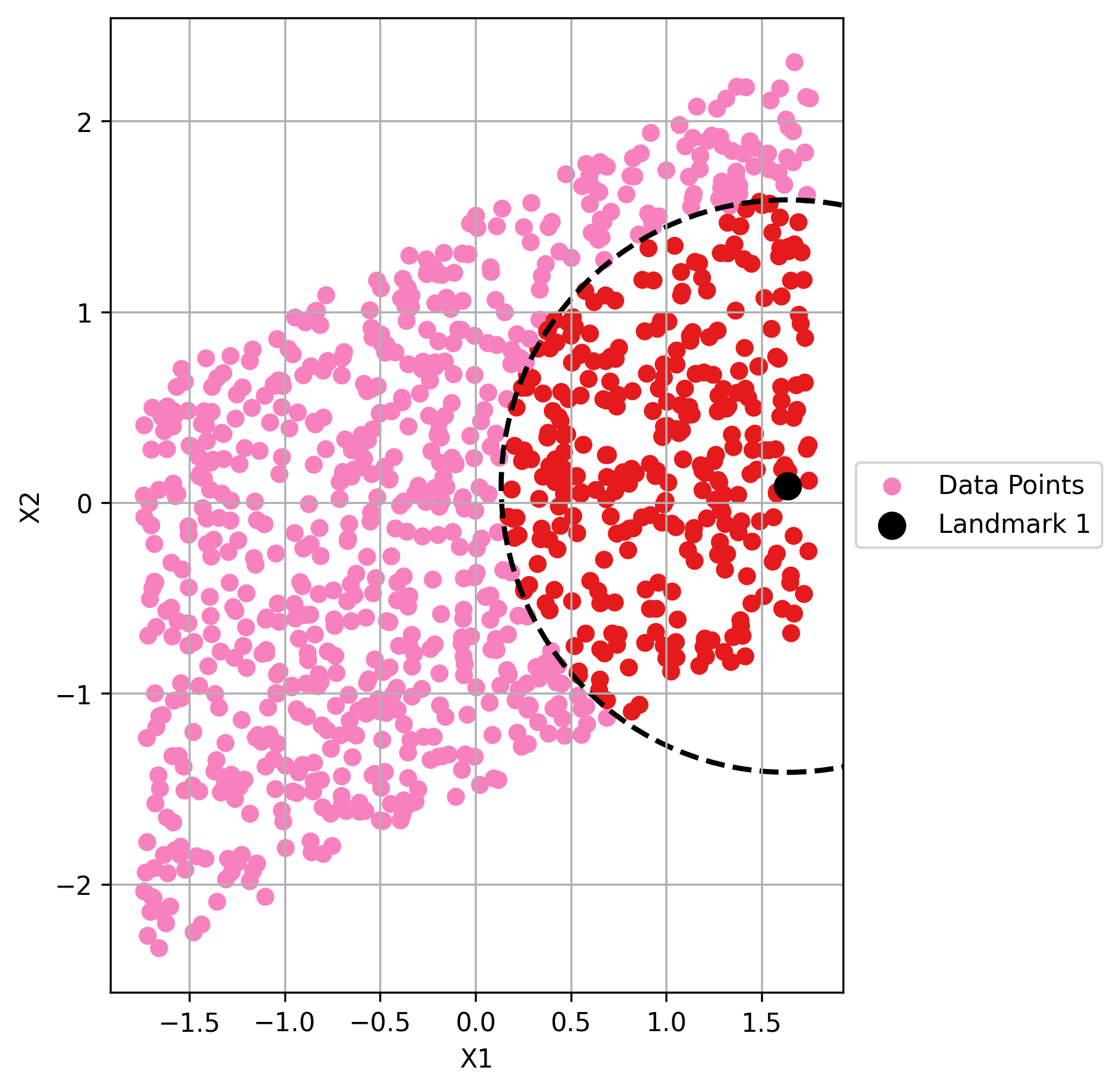}&
			\includegraphics[width=5cm]{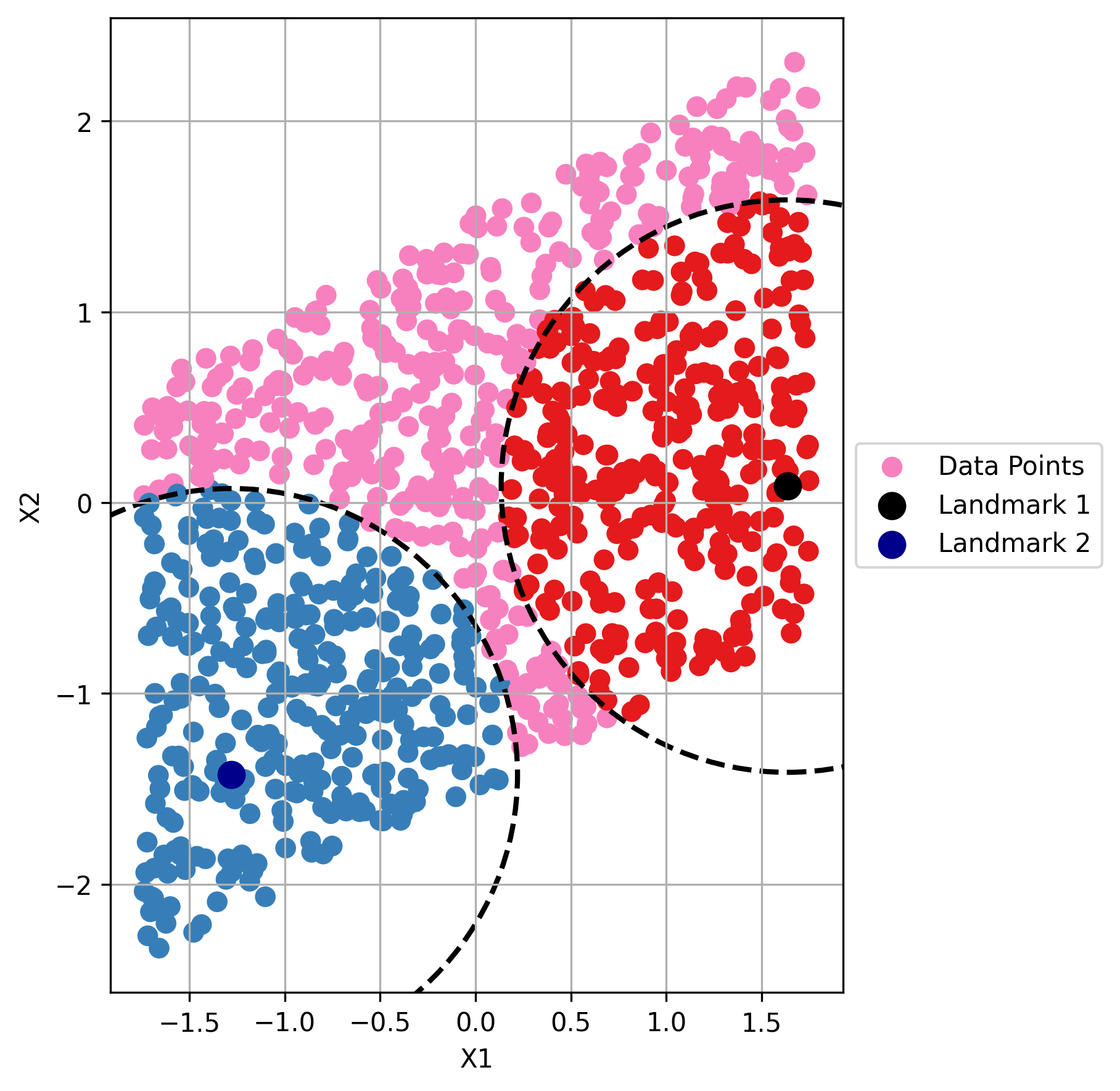}&
			\includegraphics[width=5cm]{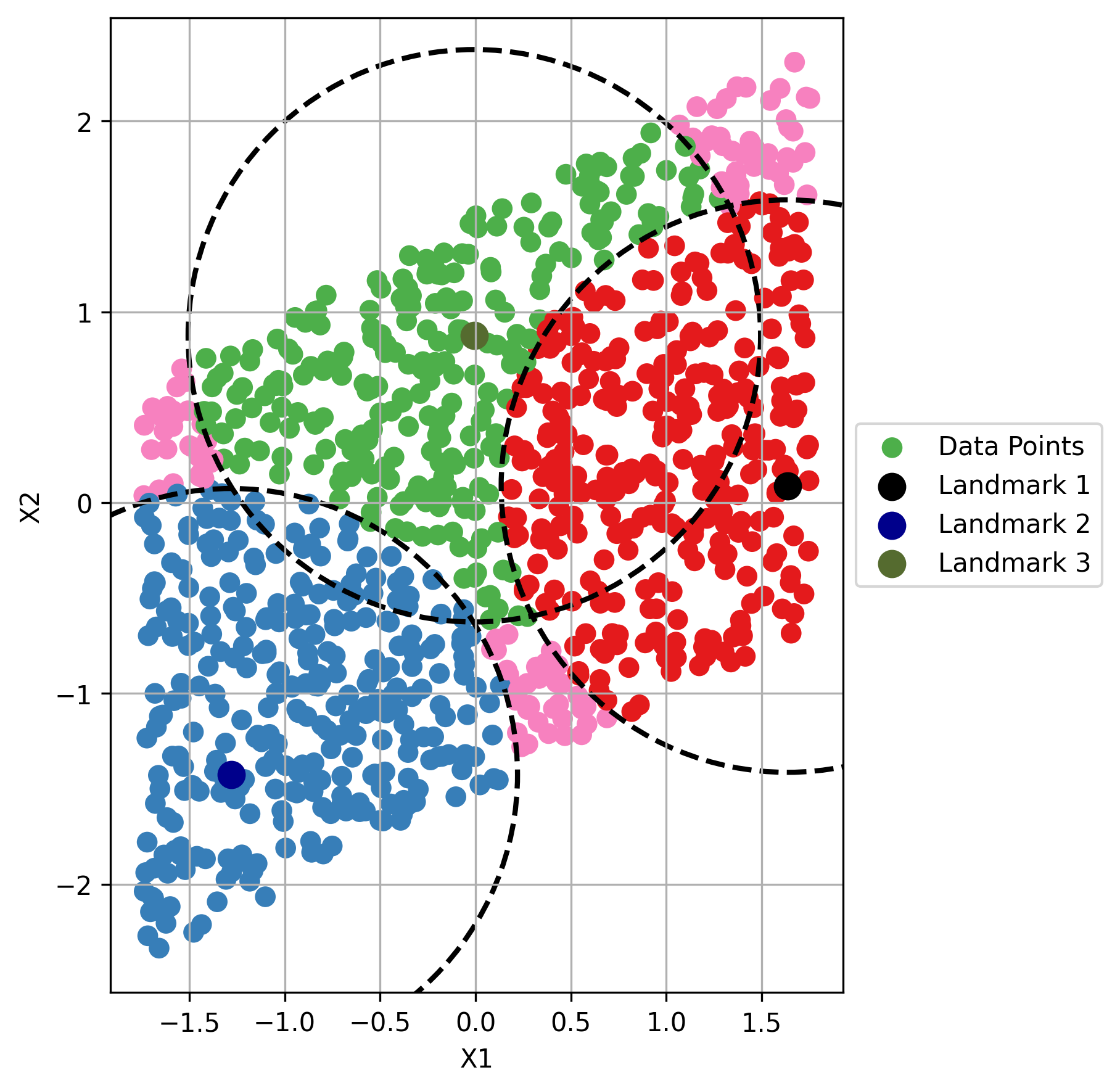}\\
			(a) Landmark 1 & (b) Landmark 2 & (c) Landmark 3 \\
			
			\includegraphics[width=5cm]{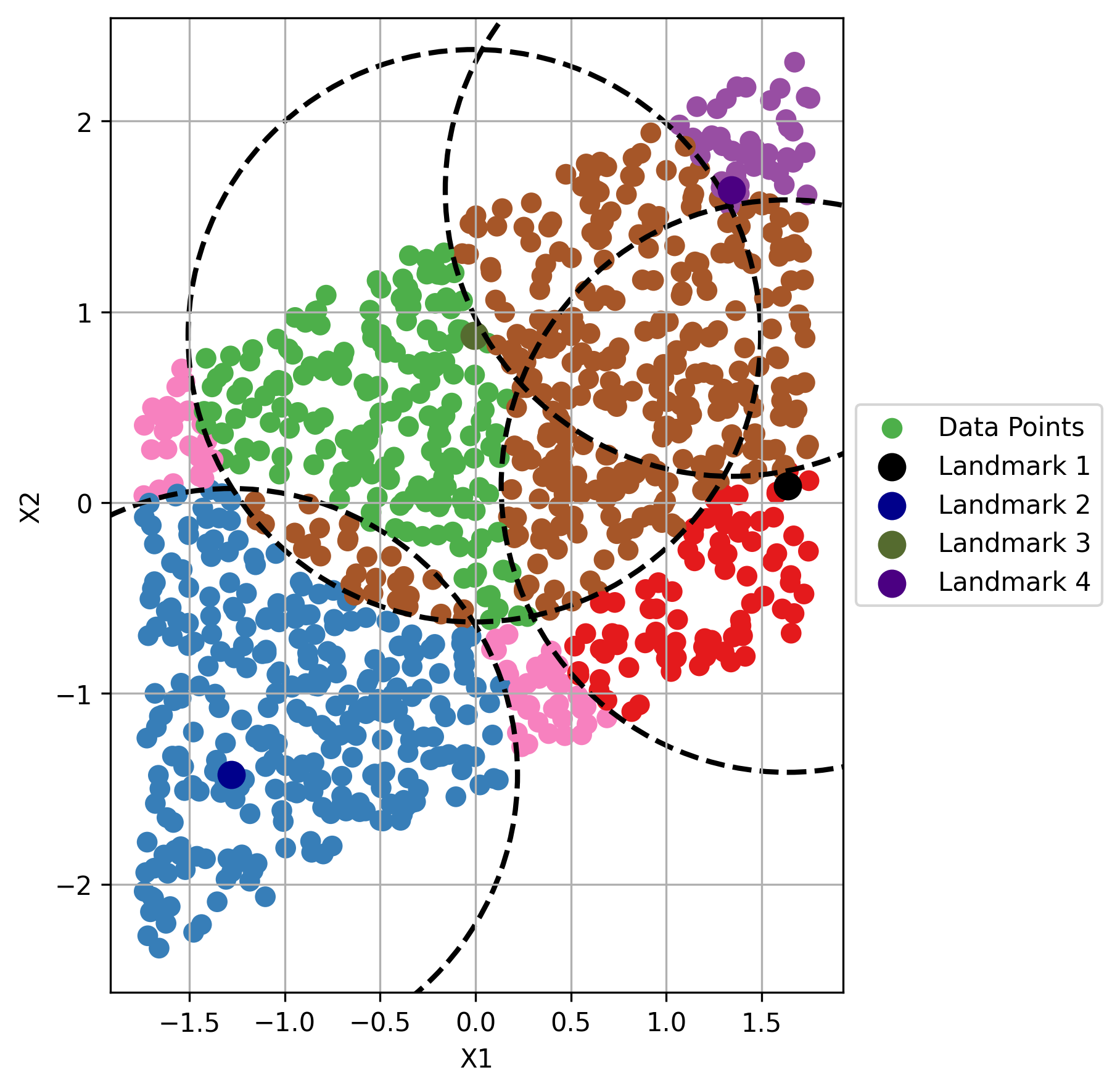}&
			\includegraphics[width=5cm]{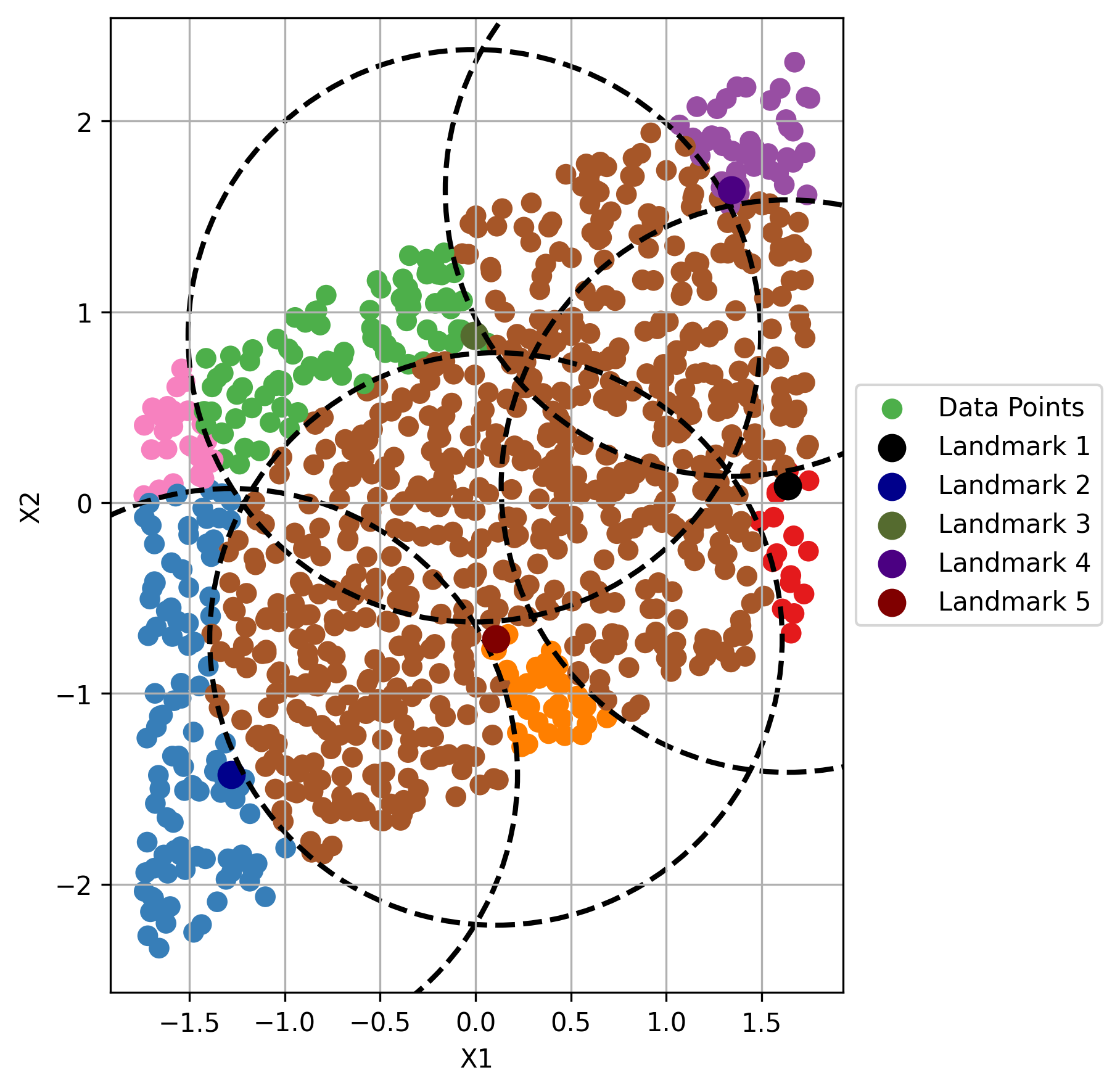}&
			\includegraphics[width=5cm]{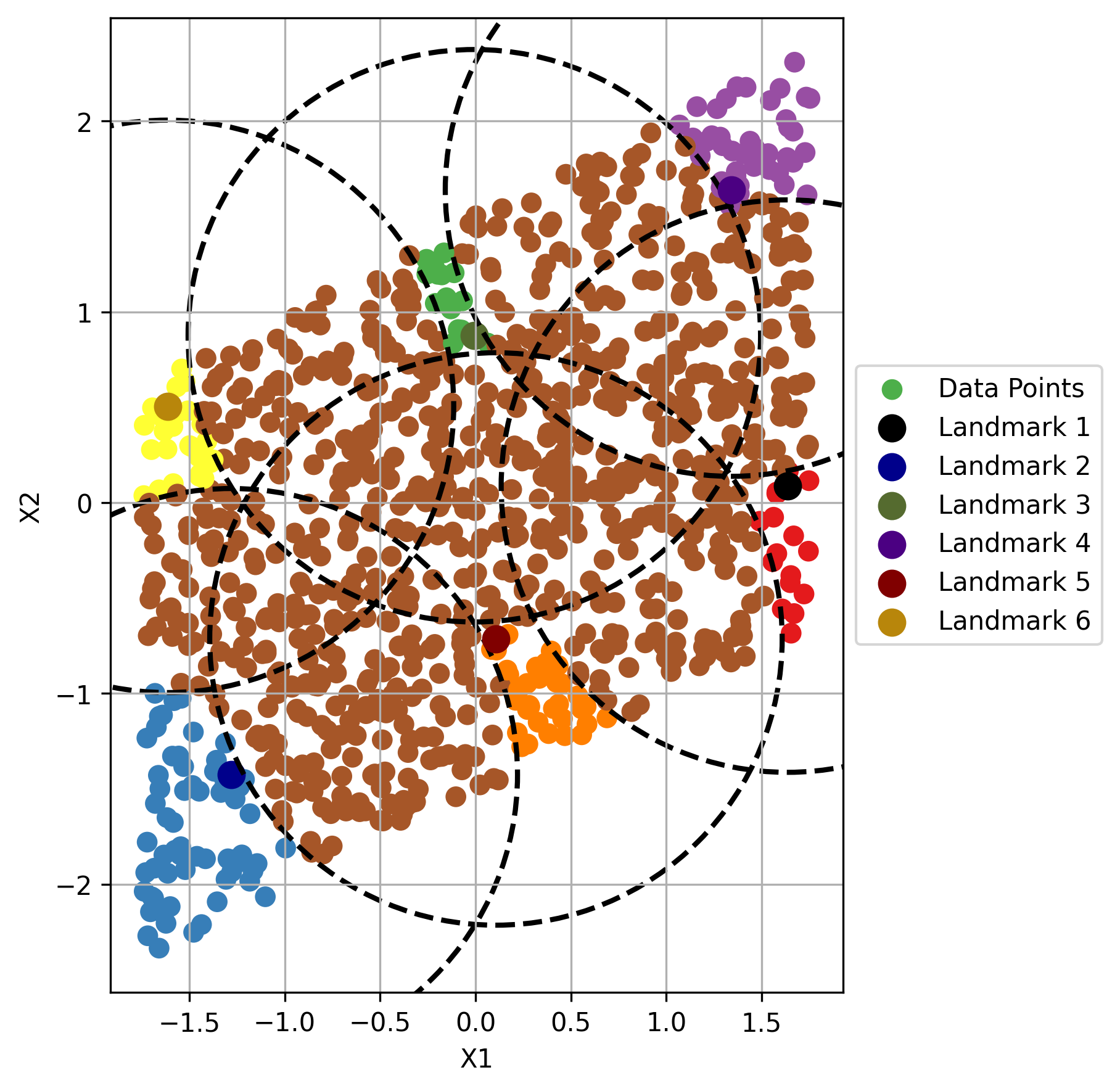}\\
			(d) Landmark 4 & (e) Landmark 5 &  (f) Landmark 6\\
		\end{tabular}
	\end{center}
	\raggedright
	\footnotesize{Notes: Figures represent the stepwise construction of the TDABM style plot. Landmarks refer to the points that would be the landmarks of the balls in a TDABM plot. Numbering of landmarks is the order of selection. Landmarks are selected from the uncovered points (pink) until all points can be found within at least one circle (all other colors). Dashed circles represent the boundaries of the balls around each landmark. Dataset 2 has a correlation between $X_1$ and $X_2$ of 0.497. Dataset 2 contains 1000 points initially drawn at random from $U \left[0,1\right]$. Standardization is applied and Dataset 2 transformed to obtain the desired correlation of approximately 0.5.}
\end{figure}

The second landmark, $l_2$ is chosen at random from all of the points which are not covered by Ball 1. A ball of radius $\epsilon$ is then constructed around $l_2$, becoming Ball 2, $B_2(X,\epsilon)$. Collectively, $B_1(X,\epsilon)$ and  $B_2(X,\epsilon)$ combine to become the cover $B(X,\epsilon)$. Points outside $B(X,\epsilon)$ are still uncovered. A third landmark is selected from $B(X,\epsilon)'$ and becomes $l_3$. The process continues until $B(X,\epsilon)' = \emptyset$, that is there are no uncovered points. Figure \ref{fig:data1t} shows the construction of the cover of Dataset 1. Figure \ref{fig:data2t} presents the process for Dataset 2.

Figure \ref{fig:data1t} shows Ball 1 being constructed in panel (a). The red points are covered by Ball 1, whilst the pink points are uncovered. A second ball is added in panel (b). There is an intersection between the two balls. The intersection points are colored in brown. Ball 2 is colored blue. The landmarks of Balls 1 and 2, $l_1$ and $l_2$ respectively, are illustrated as larger points in darker colors. Continuing the construction process, panel (c) shows the addition of Ball 3 around $l_3$. The majority of the points are covered in the example, but there are three small areas of pink on the plot. Progressively panels (d), (e) and (f), show Balls 4, 5 and 6 covering those pink points. At the point of reaching panel (f) the cover $B(X,\epsilon)$ is complete.

Within the process of replicating the theoretical TDABM algorithm in Python all of the code follows a similar pattern. Box \ref{box:code2} provides an example of the code needed to produce Ball 5. Within the accompanying code file, the use of 1 and 2 refers to the dataset. Distinction is also drawn between the two methods of constructing the example cover. Here, the use of \texttt{random\_point105} informs that landmark 5 is being selected for dataset 1. Since the code is designed as an illustration the population of the \texttt{circle\_membership} variable is abridged. Points that are only in a single ball are identified by requiring that only one of the \texttt{inside\_circleXX} dummies be equal to 1. To be in an intersection between two balls, a point must have a value of 1 for two, or more, of the \texttt{inside\_circleXX} dummies. Hence all pairwise comparisons are made. Finally a value of 8 is assigned where all of the \texttt{inside\_circleXX} dummies are 0. As the number of balls increases, so the process of assigning the \texttt{circle\_membership} values becomes longer. In the TDABM algorithm there is no need to look at the individual steps. The code provided here is purely to explain intuitively what is being done by the Python function.

Across the 6 panels of Figure \ref{fig:data1t}, the number of lines in the code increases greatly. As more balls are added, the number of lines to identify members of an individual ball increases. The number of pairwise combinations of balls is also rapidly increasing when more balls are added to the cover. The length of the code, as entered into the .ipynb file, continues to grow as the number of landmarks increases. The resulting columns in \texttt{df1} replicate the information that is held within the TDABM algorithm, and merge that information with the genuine data point information.
	
\begin{mybox}[label=box:code2]{Ball 5 Construction}
	Uncovered points are assigned a value of 8 in \texttt{circle\_membership} and so first subset to identify those points
	\begin{lstlisting}[language=Python]
		not_in_any_circle1 = df1[df1['circle_membership'] == 8]
	\end{lstlisting}
	Selection of the landmark now just requires identifying the lowest index value and finding that point in the data. The landmark is named \texttt{random\_point105}.
	\begin{lstlisting}[language=Python]
		fifth_random_index1 = np.random.choice(not_in_any_circle1.index)
		random_point105 = df1.loc[fifth_random_index1]
	\end{lstlisting}
	Identify the distance of each data point from the selected landmark, \texttt{random\_point105}.
	\begin{lstlisting}[language=Python]
		df1['distance_fifth_circle'] = np.sqrt((df1['X1'] - random_point105['X1'])**2 + (df1['X2'] - random_point105['X2'])**2)
	\end{lstlisting}
	Find the points in \texttt{df1} with distance less than the ball radius. In this example the radius is 1.5. \texttt{inside\_circle05} denotes that a point is inside the circle (ball) around landmark 5.
	\begin{lstlisting}[language=Python]
		radius = 1.5
		df1['inside_circle05'] = (df1['distance_fifth_circle'] <= radius).astype(int)
	\end{lstlisting}
	Finally, a membership variable is constructed. This will take the value of the ball number if the point is only in one ball. Will progressively take the value 7 when the point is in two balls and 8 if the point is not in any balls. The example has 6 balls, hence the use of 7 and 8 as the next two integers.
	\begin{lstlisting}[language=Python]
		df1.loc[(df1['inside_circle01'] == 1) & (df1['inside_circle02'] == 0) & (df1['inside_circle03'] == 0) & (df1['inside_circle04'] == 0) & (df1['inside_circle05'] == 0), 'circle_membership'] = 1
	\end{lstlisting}
	All of the pairwise combinations are allocated 7 if both have the value 1 for the \texttt{inside\_circle}
	\begin{lstlisting}[language=Python]
		df1.loc[(df1['inside_circle01'] == 1) & (df1['inside_circle02'] == 1), 'circle_membership'] = 7
	\end{lstlisting}
	A final line assigns the value 8 to any point which is not in any ball
	\begin{lstlisting}[language=Python]
		df1.loc[(df1['inside_circle01'] == 0) & (df1['inside_circle02'] == 0) & (df1['inside_circle03'] == 0) & (df1['inside_circle04'] == 0) & (df1['inside_circle05'] == 0), 'circle_membership'] = 8
	\end{lstlisting}
\end{mybox}

The process can be replicated on Dataset 2. The case of Dataset 2 is plotted in Figure \ref{fig:data2t}. In the construction of Ball 2, panel (b) of Figure \ref{fig:data2t}, there is no intersection between the first two balls. Only when Ball 3 is added in panel (c) are any overlaps noted. There are three pockets of uncovered points. Panels (d) to (f) add the balls required to complete the cover of Dataset 2. 

Figures \ref{fig:data1t} and \ref{fig:data2t} show how a cover is constructed. The cover $B(X,\epsilon)$ is the first stage of the TDABM algorithm. The second stage is to construct the actual map of the data. The visualization is created by converting the information amongst the ball members into a representation of each ball. Each ball is shown in a TDABM graph as a disc. The number of points in the ball informs on the density of the data within the neighborhood of the landmark. In a TDABM graph, the number of points in the ball is captured in the size of the disc that represents each ball. There is also information about $Y$ for each data point. To preserve information about $Y$, an average across all ball members is taken. The average value of $Y$ informs the color of each disc. In order to understand the relative positioning of the balls, edges are drawn between any pairs of balls for which the intersection is non-empty. From Figures \ref{fig:data1t} and \ref{fig:data2t}, it can be seen that there are many non-empty intersections. 

\begin{figure}
	\begin{center}
		\caption{Towards the TDABM Style Graph: Random}
		\label{fig:tbm1}
		\begin{tabular}{c c}
			\includegraphics[width=6cm,height=6cm]{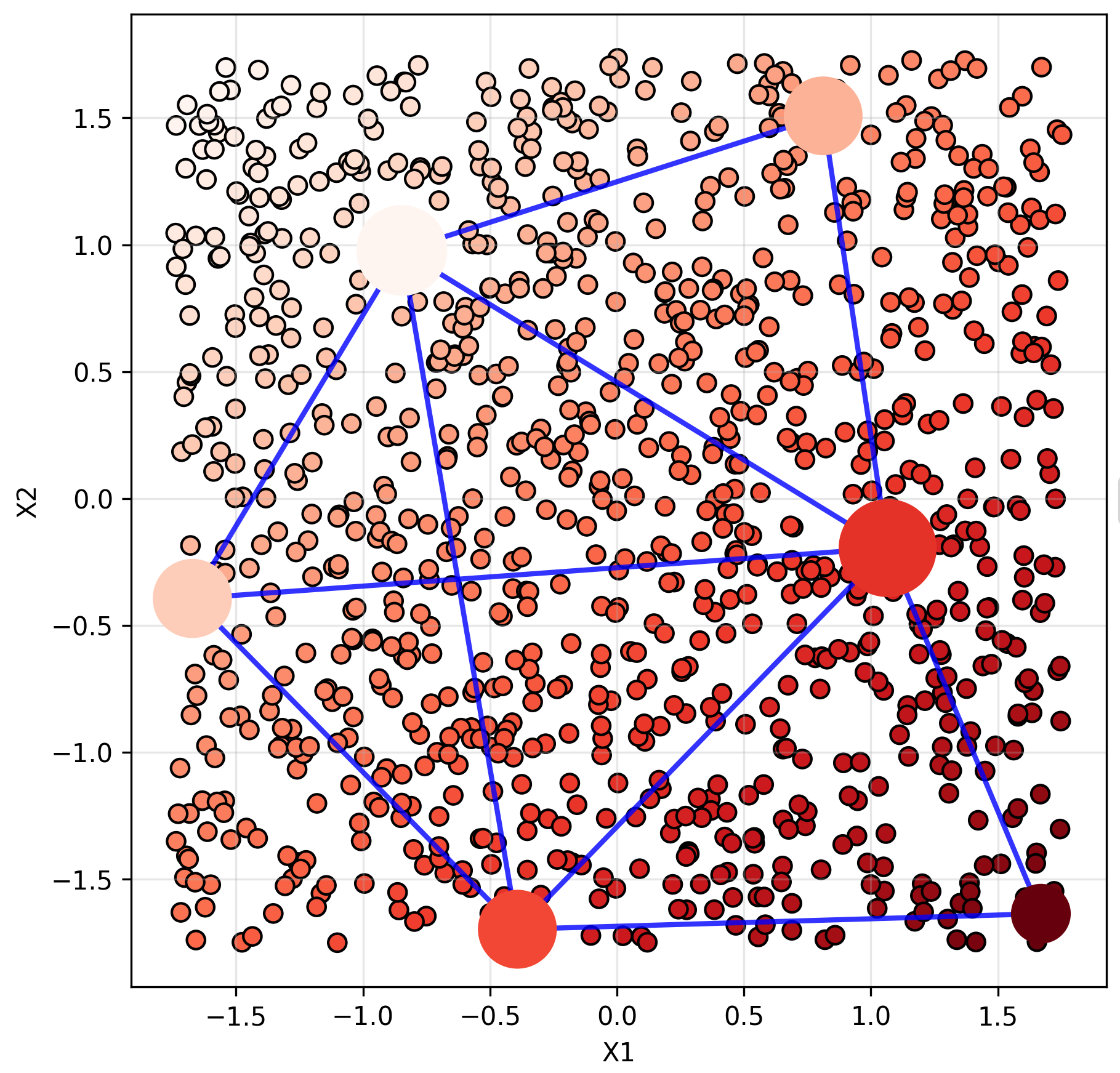}&
			\includegraphics[width=6cm,height=6cm]{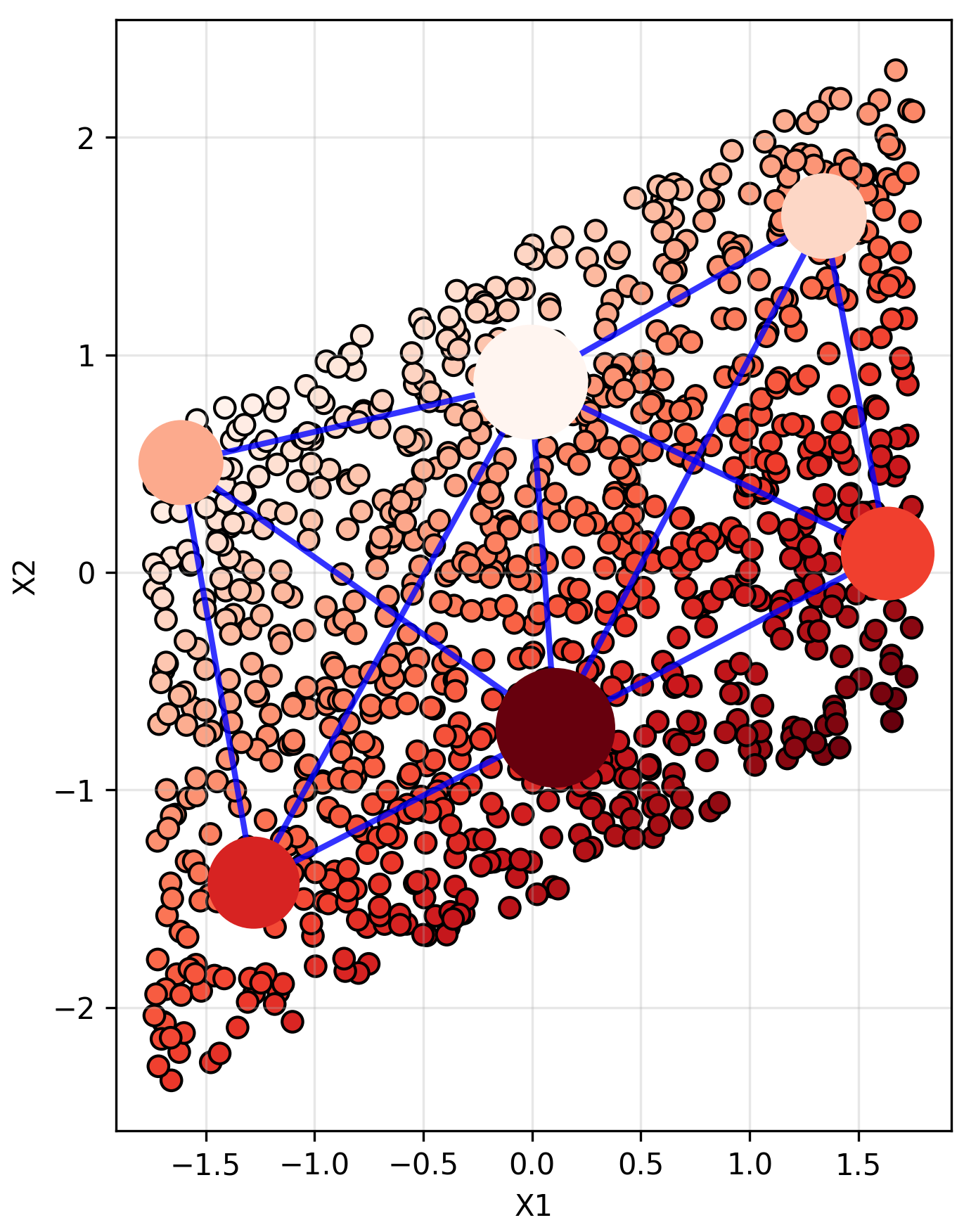}\\
			(a) Dataset 1 with Discs  & (b) Dataset 2 with Discs \\
			\includegraphics[width=7cm]{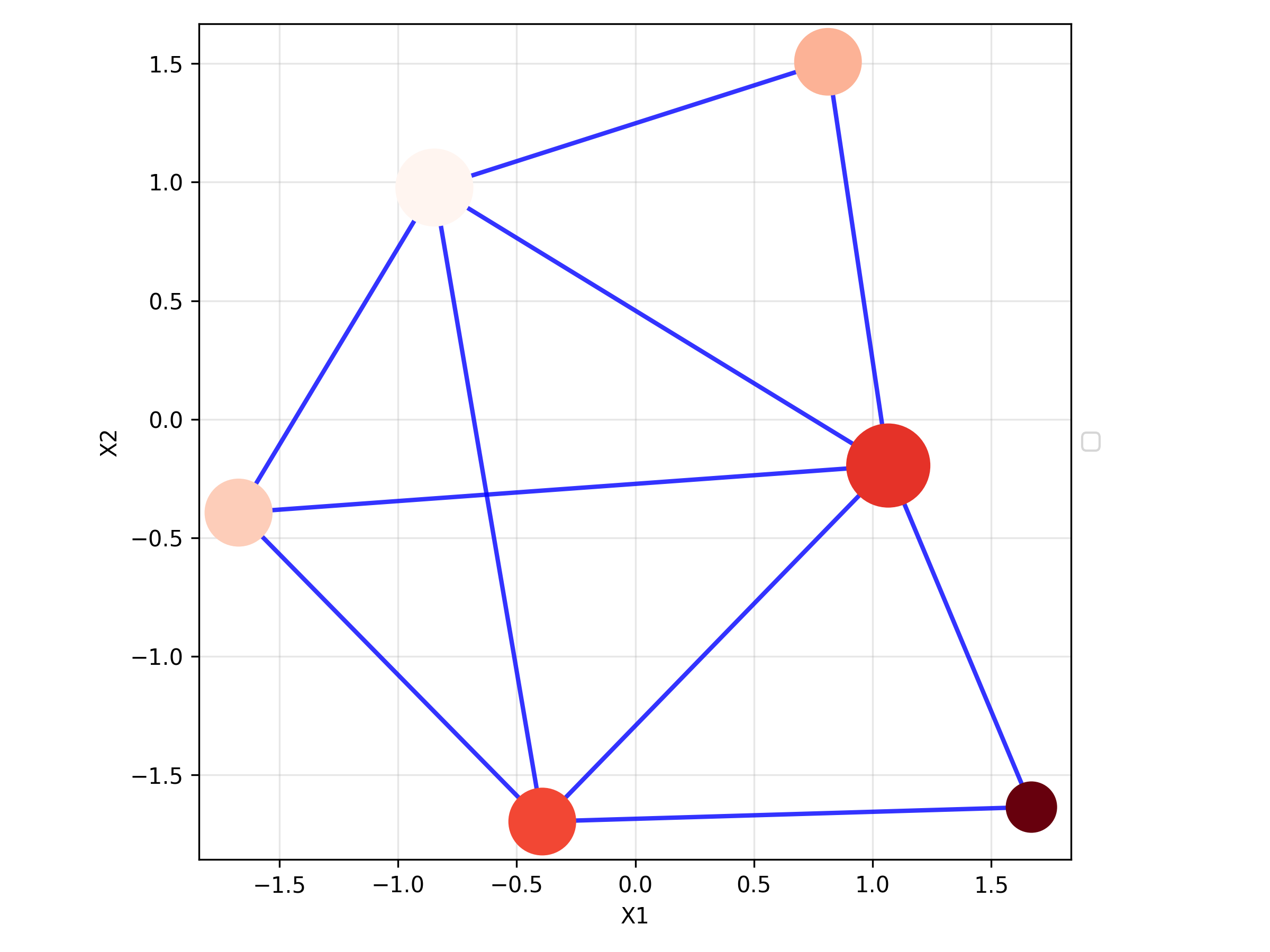}&
			\includegraphics[width=7cm]{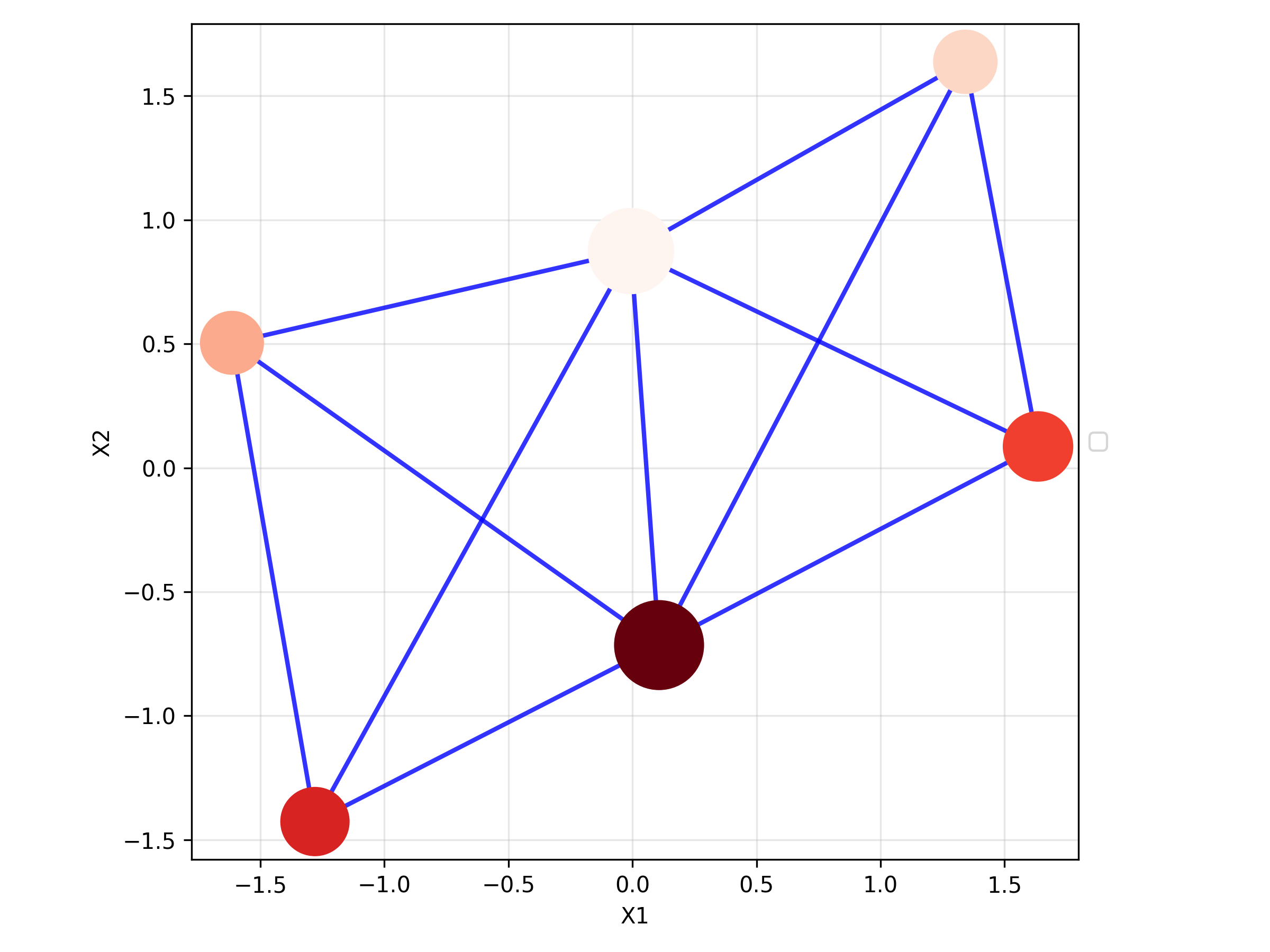}\\
			(c) Structure with Axes 1 & (b) Structure with Axes 2 \\
			\includegraphics[width=7cm]{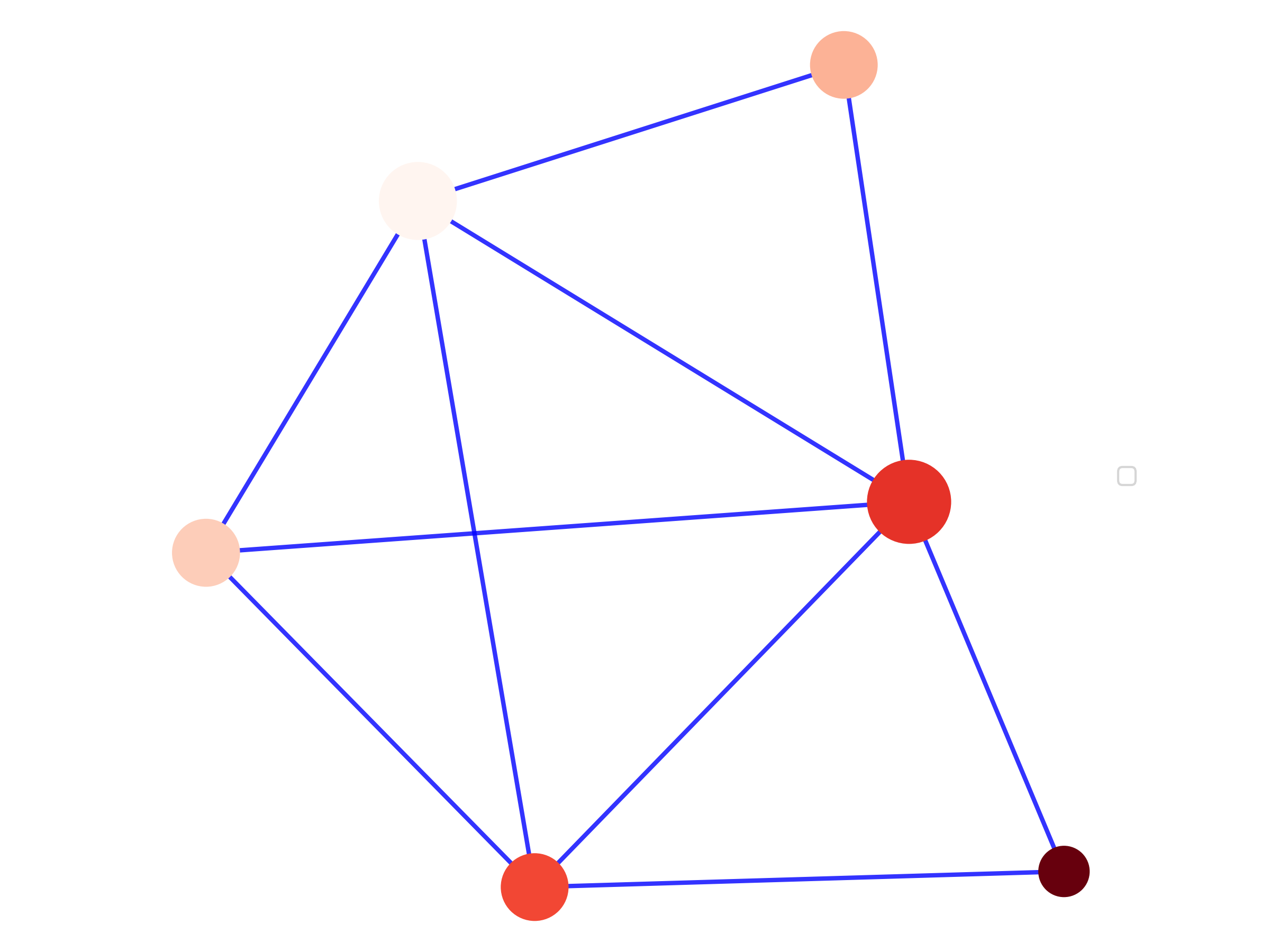}&
			\includegraphics[width=7cm]{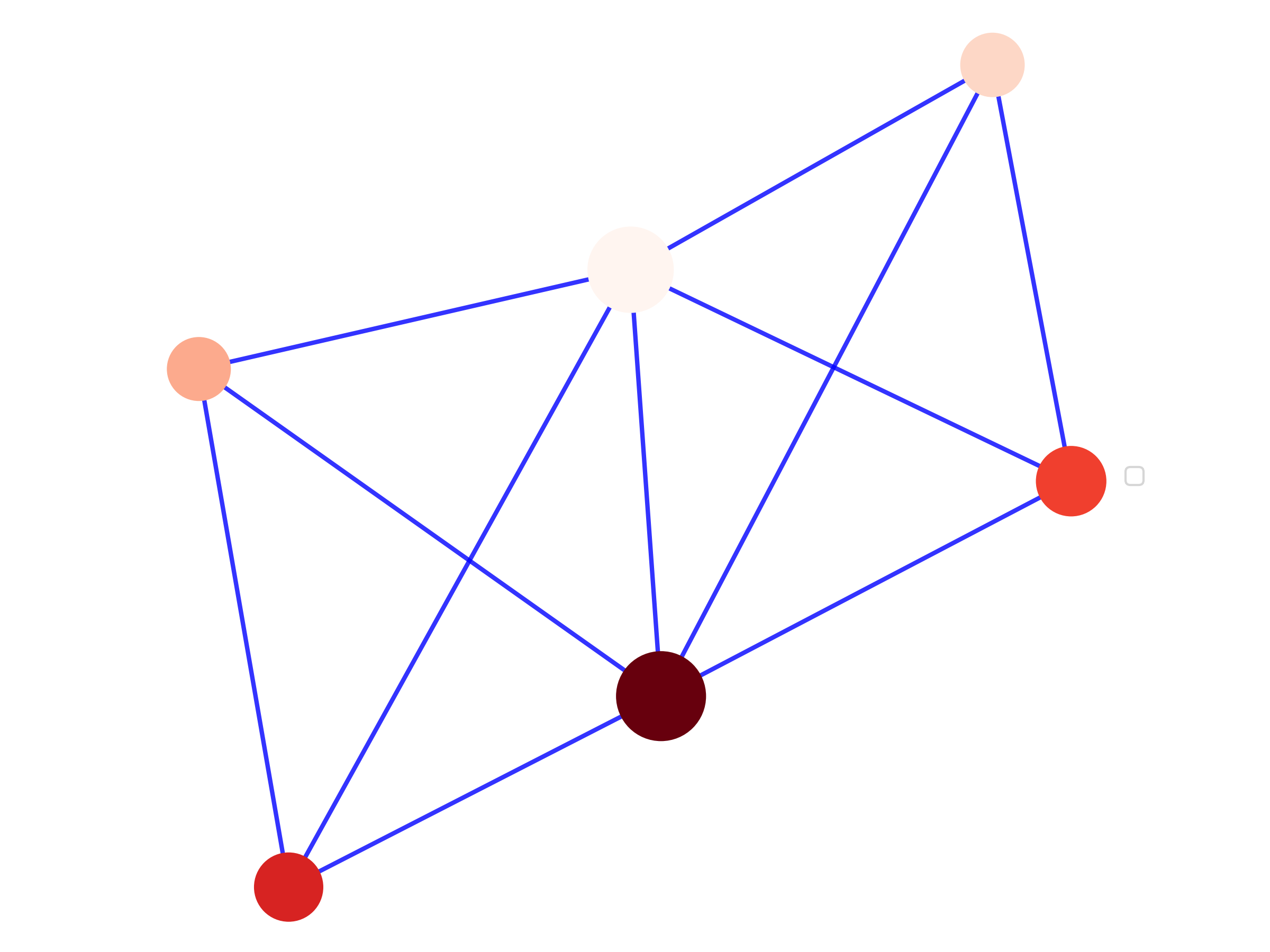}\\
			(a) TDABM Equivalent 1 & (b) TDABM Equivalent 2 \\
		\end{tabular}
	\end{center}
	\raggedright
	\footnotesize{Notes: Figures represent the transition from the full dataset to the TDABM style plot. Dataset 1 has no correlation between $X_1$ and $X_2$, whilst Dataset 2 has a correlation between $X_1$ and $X_2$ of 0.497. Both datasets contain 1000 points initially drawn at random from $U \left[0,1\right]$. Standardization is applied and Dataset 2 transformed to obtain the desired correlation of approximately 0.5.}
\end{figure}

The construction of the TDABM graphs for the Dataset 1 and Dataset 2 is provided in Figure \ref{fig:tbm1}. In panels (a) and (b), the original datapoints are visible. Coloration of the data points is according to the value of $Y$. The balls are shown by the dashed circles, as they were in Figures \ref{fig:data1t} and \ref{fig:data2t}. All of the information that would be captured in the TDABM graph is visible in panels (a) and (b). The TDABM style graph is overlaid onto the plot. Discs are sized according to the number of points within the ball. Discs are colored according to the average value of $Y$ across all of the points within the ball. The edges are added across all intersections that are non-empty.

Panels (c) and (d) show the removal of the individual datapoints. The axes are still shown, and hence the balls remain in the location of their landmarks. The plot is less cluttered and can still be seen to hold all of the information from the data. The final step that can be made with the theoretical example is to remove the axes to leave a random representation. Panels (e) and (f) show the final theoretical representation. Because the balls can be in any number of dimensions, the TDABM graph is abstract. Only in the two-dimensional case can information about the location of the landmarks be maintained in the plot. In Python, the \texttt{PyBallMapper} function does not preserve the co-ordinates of the landmarks, meaning that even representations of two-dimensional data are abstract. 

The TDABM algorithm can thus be summarized as creating a cover $B(X,\epsilon)$ from the sequential selection of landmark points $l_b$ from the set of uncovered points by drawing balls of radius $\epsilon$ around the selected landmark. Iteratively selecting landmarks continues until all points are in at least one ball. The TDABM graph is constructed from $B(X,\epsilon)$ using discs which are sized according to the number of points in each ball and colored according to a function on the points in the ball. Edges in a TDABM graph appear between any pair of balls which have a non-empty intersection.

\section{Ball Mapper as Implemented in Python}
\label{sec:art}

In order to allow the controlled selection of landmarks within the TDABM algorithm, the implementation provided in Python is slightly different to the implementation described in the methodology section. The \texttt{PyBallMapper} function takes the first uncovered point in the dataset as being the next landmark. For $l_1$ this means that the first row of the dataset is the landmark.  Box \ref{box:ap1} shows the difference in the code from the previous example to the recreation of the Python implementation. The result is different TDABM representations, but with the data unchanged the overall inference from the TDABM graph will be consistent. 

\begin{mybox}[label=box:ap1]{Selection of Landmark 1}
	In the Python implementation the selection of the first landmark, $l_1$, is made by:
	\begin{lstlisting}[language=Python]
		first_point_index = 0
		first_point1 = df1.loc[first_point_index]
	\end{lstlisting}
	When following the theoretical TDABM algorithm, selection of the random point that will be the first landmark is made using \texttt{numpy}
	\begin{lstlisting}[language=Python]
		random_index101 = np.random.choice(df1.index)
		random_point101 = df1.loc[random_index101]
	\end{lstlisting}
	Hence \texttt{first\_point1} is expected to be different from \texttt{random\_point101}. Only in the case where the random selection chooses the lowest index amongst the uncovered points will the selection be identical.
\end{mybox}

\begin{figure}
	\begin{center}
		\caption{Dataset 1 Sequential Cover}
		\label{fig:data1b}
		\begin{tabular}{c c c}
			\includegraphics[width=5cm]{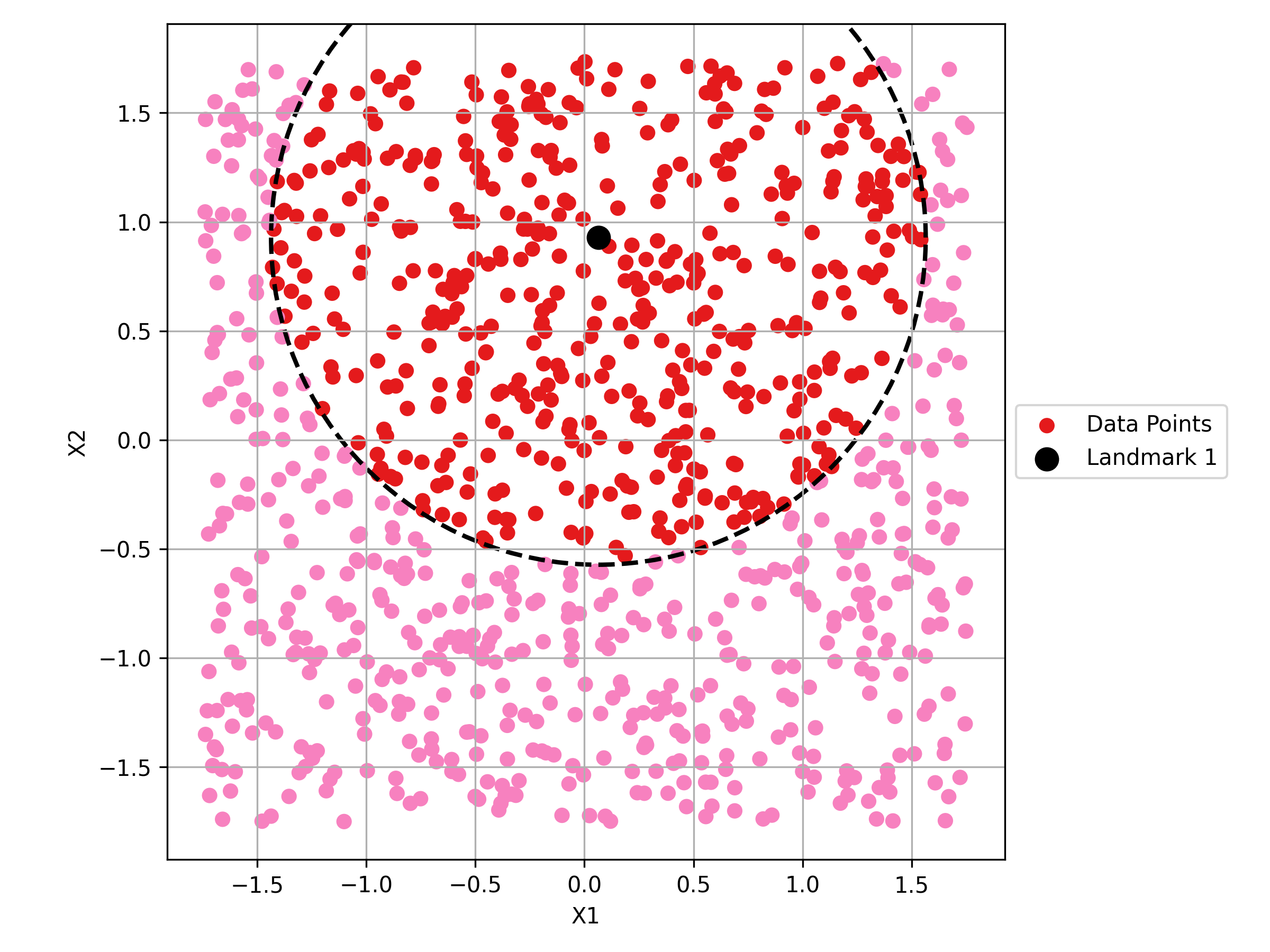}&
			\includegraphics[width=5cm]{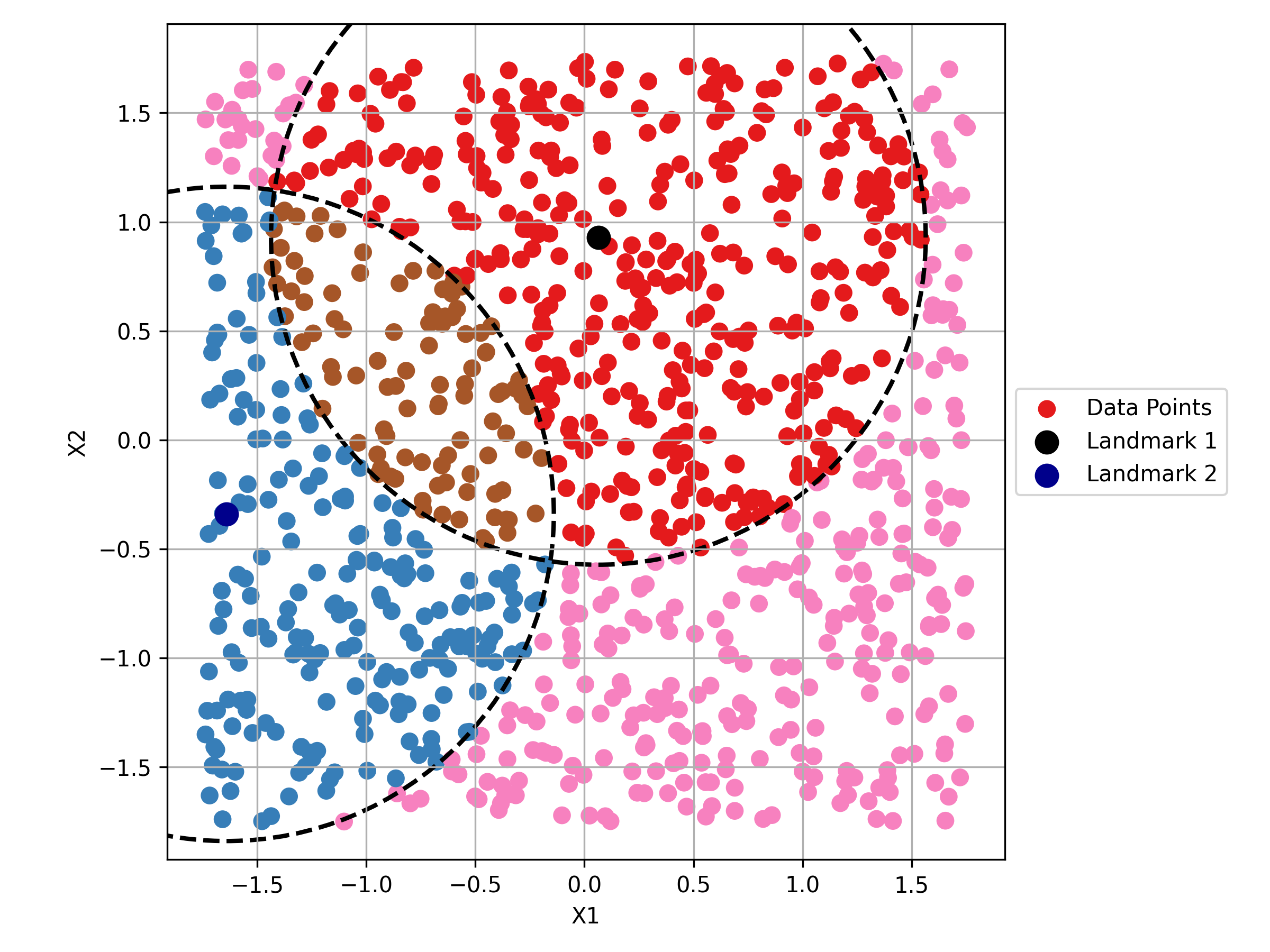}&
			\includegraphics[width=5cm]{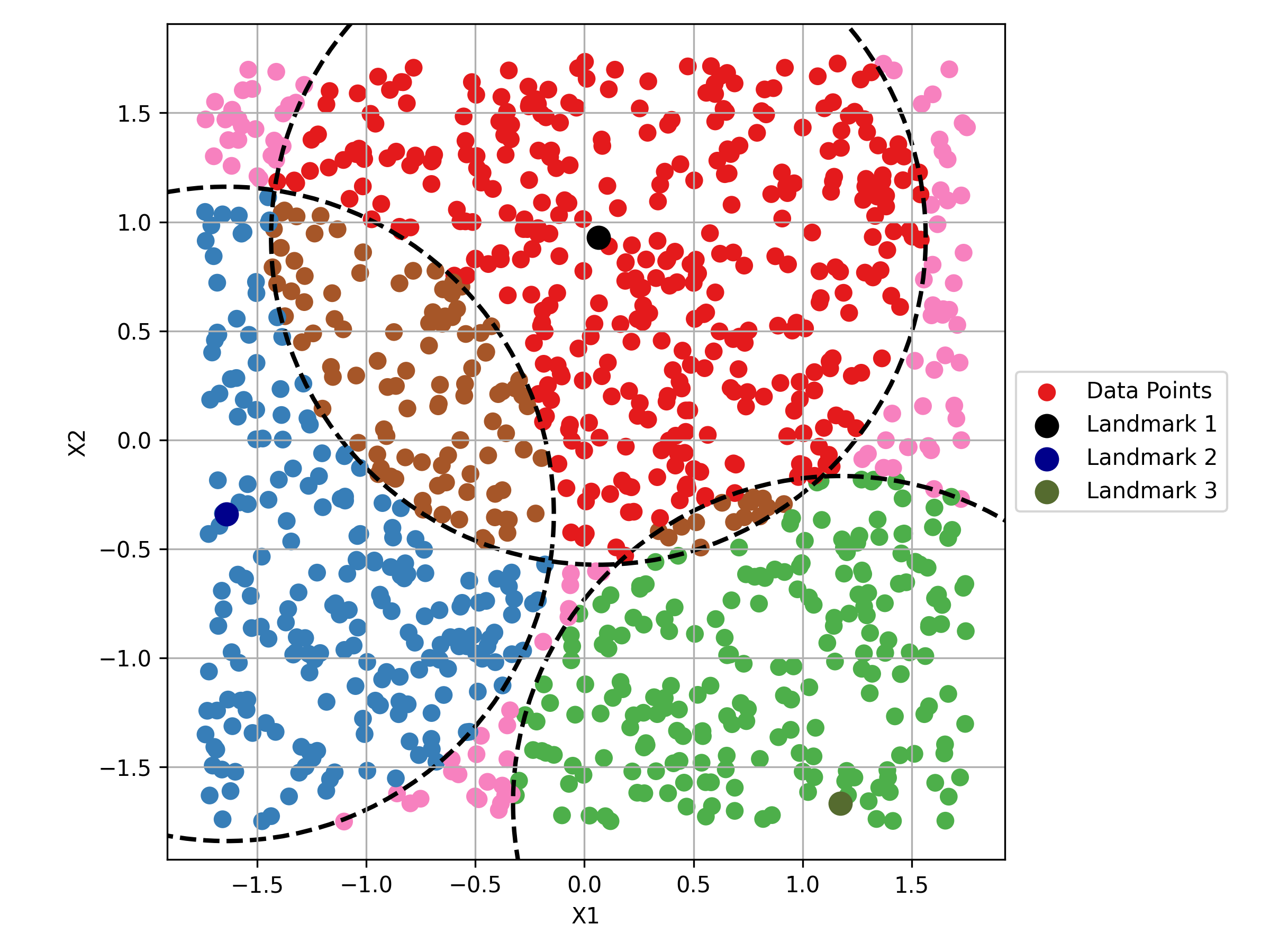}\\
			(a) Landmark 1 & (b) Landmark 2 & (c) Landmark 3  \\
		
			\includegraphics[width=5cm]{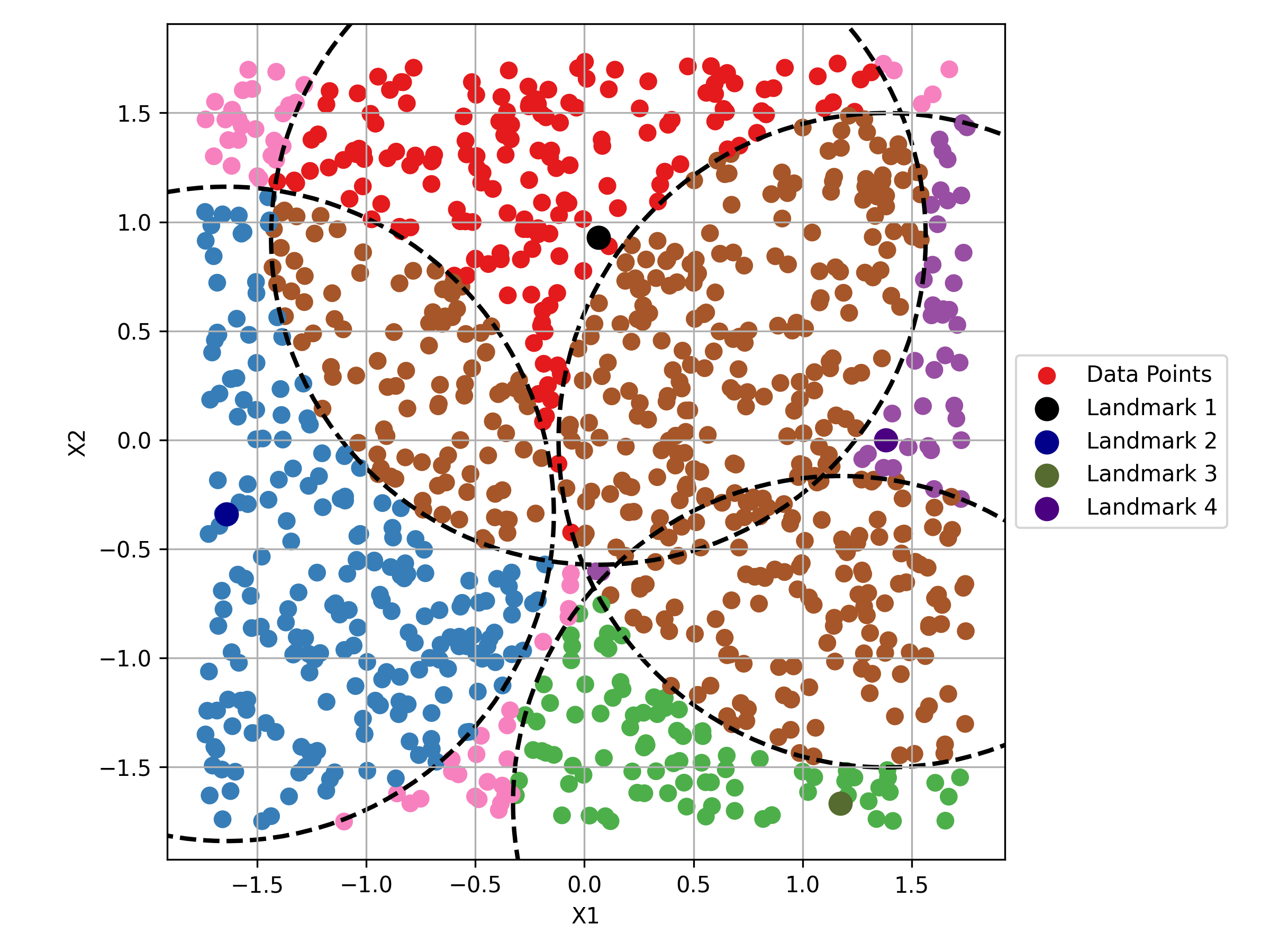} & 
			\includegraphics[width=5cm]{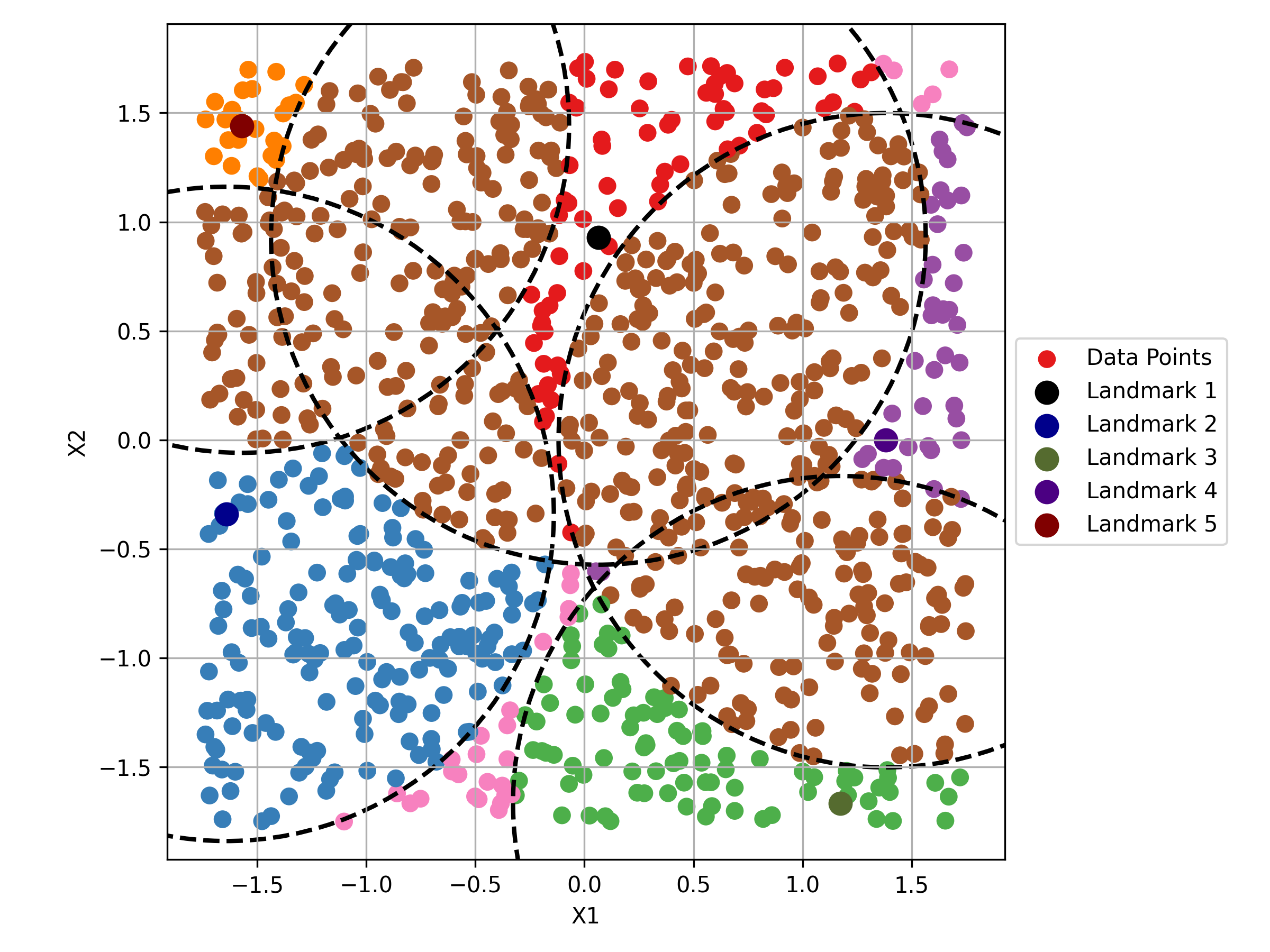}&
			\includegraphics[width=5cm]{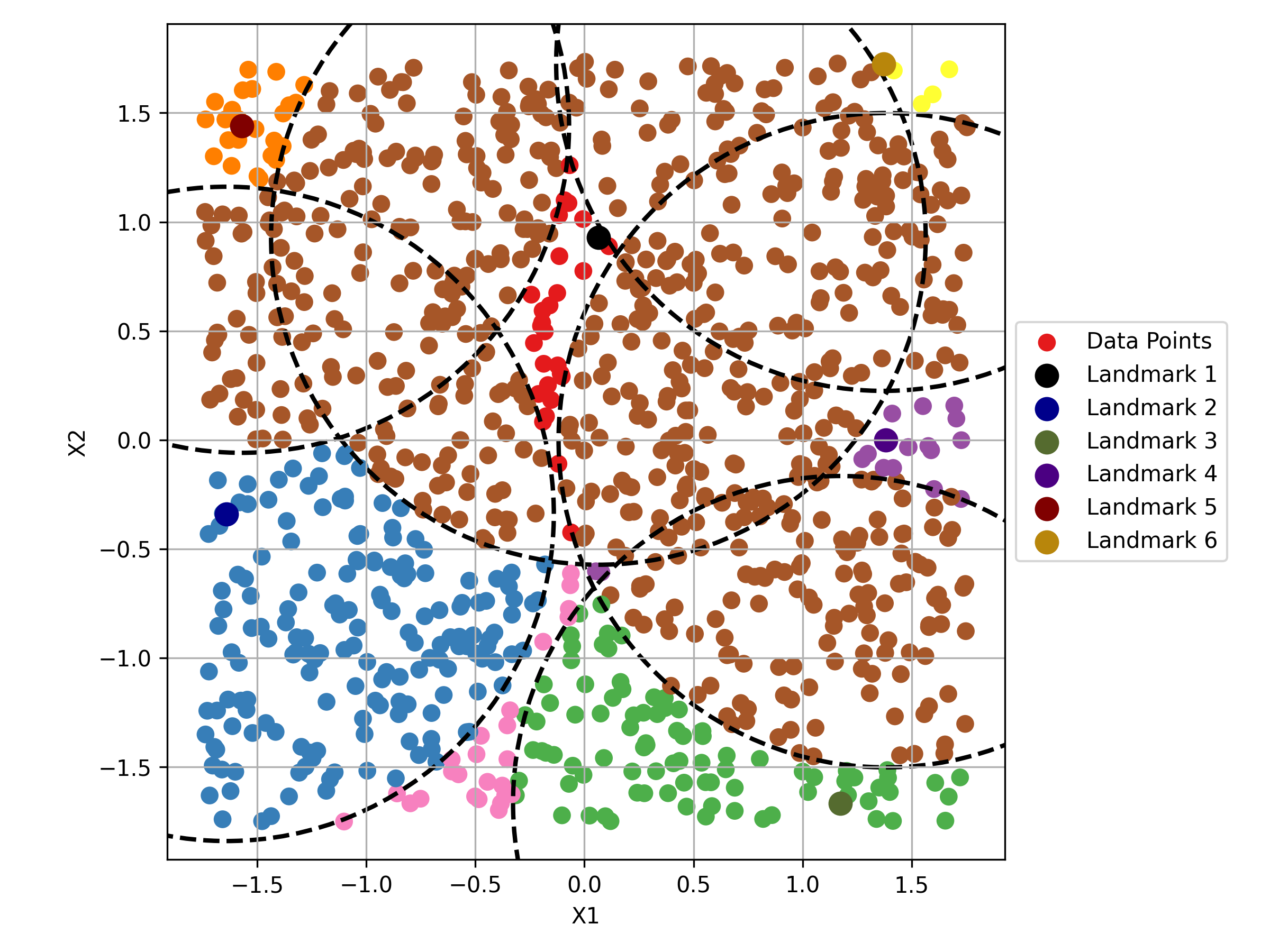}\\
			(d) Landmark 4 & (e) Landmark 5 & (f) Landmark 6\\
			&\includegraphics[width=5cm]{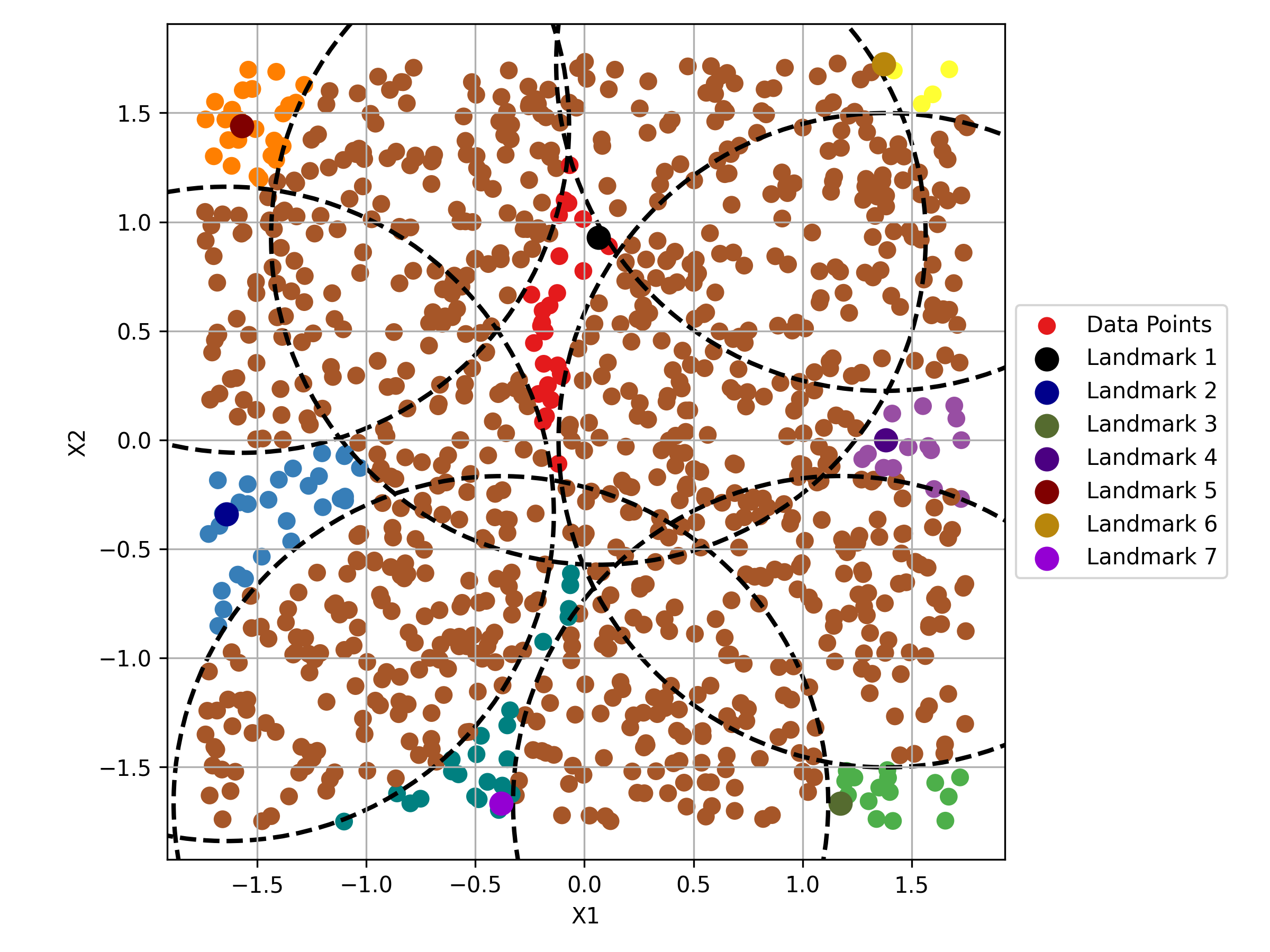}&\\
			& (g) Landmark 7 & \\
			
		\end{tabular}
	\end{center}
		\raggedright
	\footnotesize{Notes: Figures represent the stepwise construction of the TDABM style plot. Landmarks refer to the points that would be the landmarks of the balls in a TDABM plot. Numbering of landmarks is the order of selection. Landmarks are selected from the uncovered points (pink) until all points can be found within at least one circle (all other colors). Landmark selection is according the lowest index which is uncovered. Dashed circles represent the boundaries of the balls around each landmark. Dataset 1 has two variables $X_1$ and $X_2$, and contains 1000 points. $X_1$ and $X_2$ are drawn at random from $U \left[0,1\right]$. Standardization is applied to allow comparison with Dataset 2.}
\end{figure}

Having selected $l_1$, the process of constructing Ball 1 is identical. A ball of radius $\epsilon$ is drawn around $l_1$. All points within Ball 1 are covered and the remaining points are uncovered. The lowest index within the uncovered set becomes $l_2$. Figure \ref{fig:data1b} shows the construction of the cover, $B(X,\epsilon)$, when the landmarks are selected sequentially. The contrast between Figure \ref{fig:data1b} and \ref{fig:data1t} provides some indication of the variation in TDABM graphs which can occur as a result of the random selection.

An immediate observation in the comparison of Figure \ref{fig:data1b} with \ref{fig:data1t} is that there are now 7 balls required to complete the cover. The different choice of landmarks is most easily seen on the plots with the fewest balls. When a random point was selected, a point to the centre right of the plot was chosen, resulting in a large coverage of the data. However, the second point then left a small pocket of uncovered points in the extreme bottom right. In the case of sequential selection, the first point selected is towards to the top centre. The second ball then creates a pocket of uncovered points in the top right. There is some rotational symmetry to this point. Point 3 in the sequential cover is then to the lower right and leaves an area to the top right corner where there is potential for a ball that does not cover all points. Ball 3 in the theoretical example does not leave such a large uncovered area. 

\begin{figure}
	\begin{center}
		\caption{Dataset 2 Sequential Cover}
		\label{fig:data2b}
		\begin{tabular}{c c c}
			\includegraphics[width=5cm]{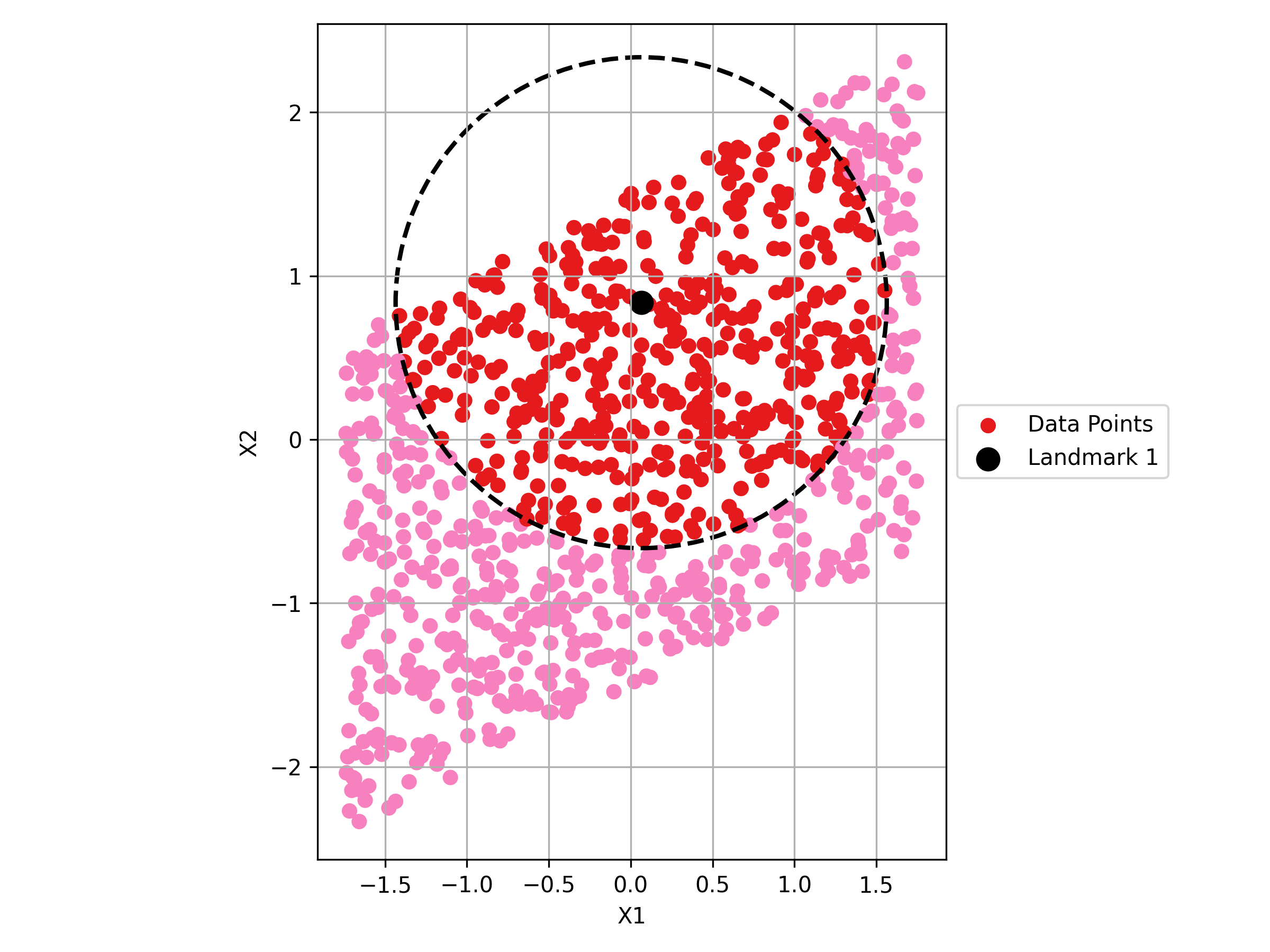}&
			\includegraphics[width=5cm]{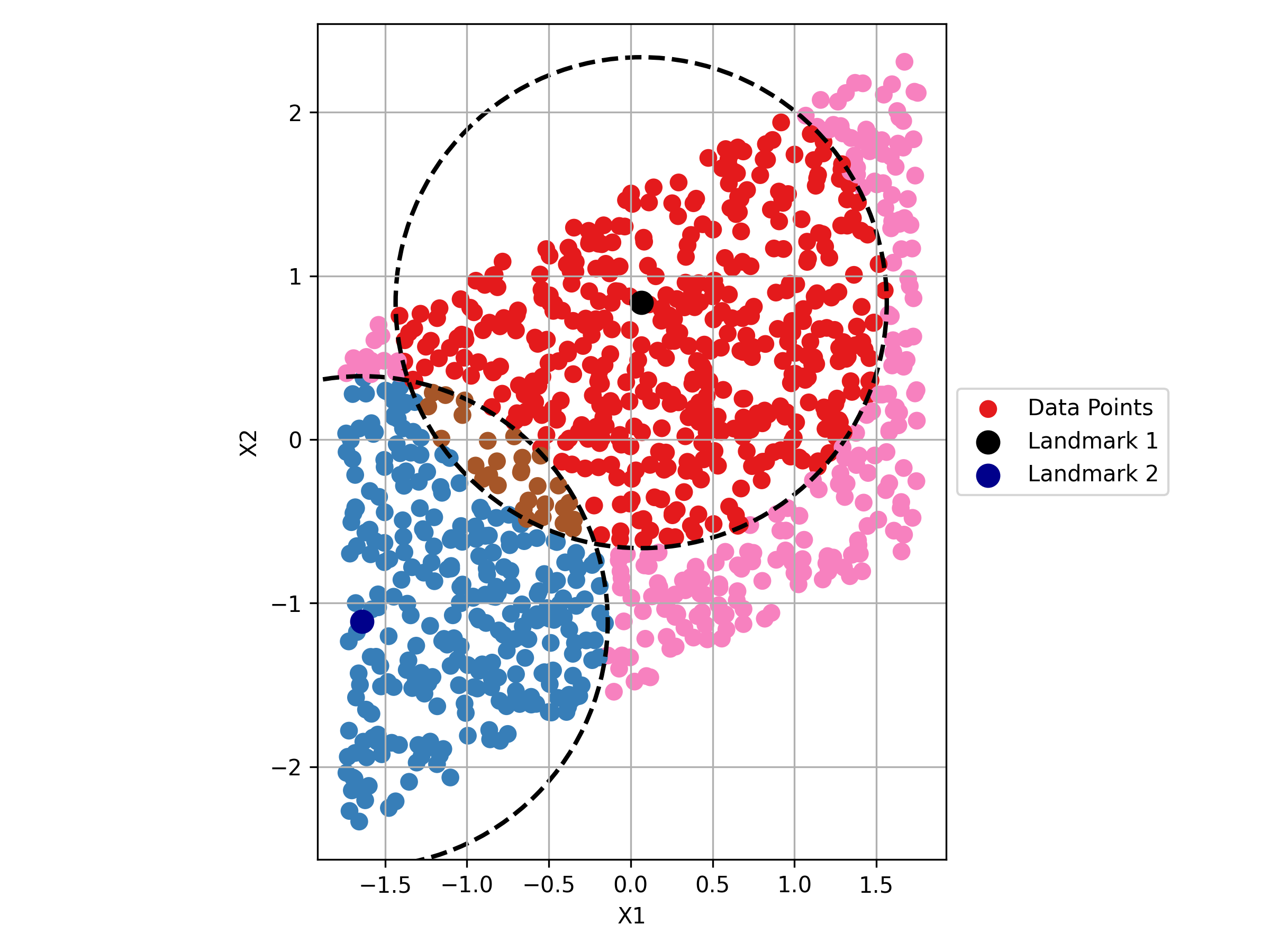}&
			\includegraphics[width=5cm]{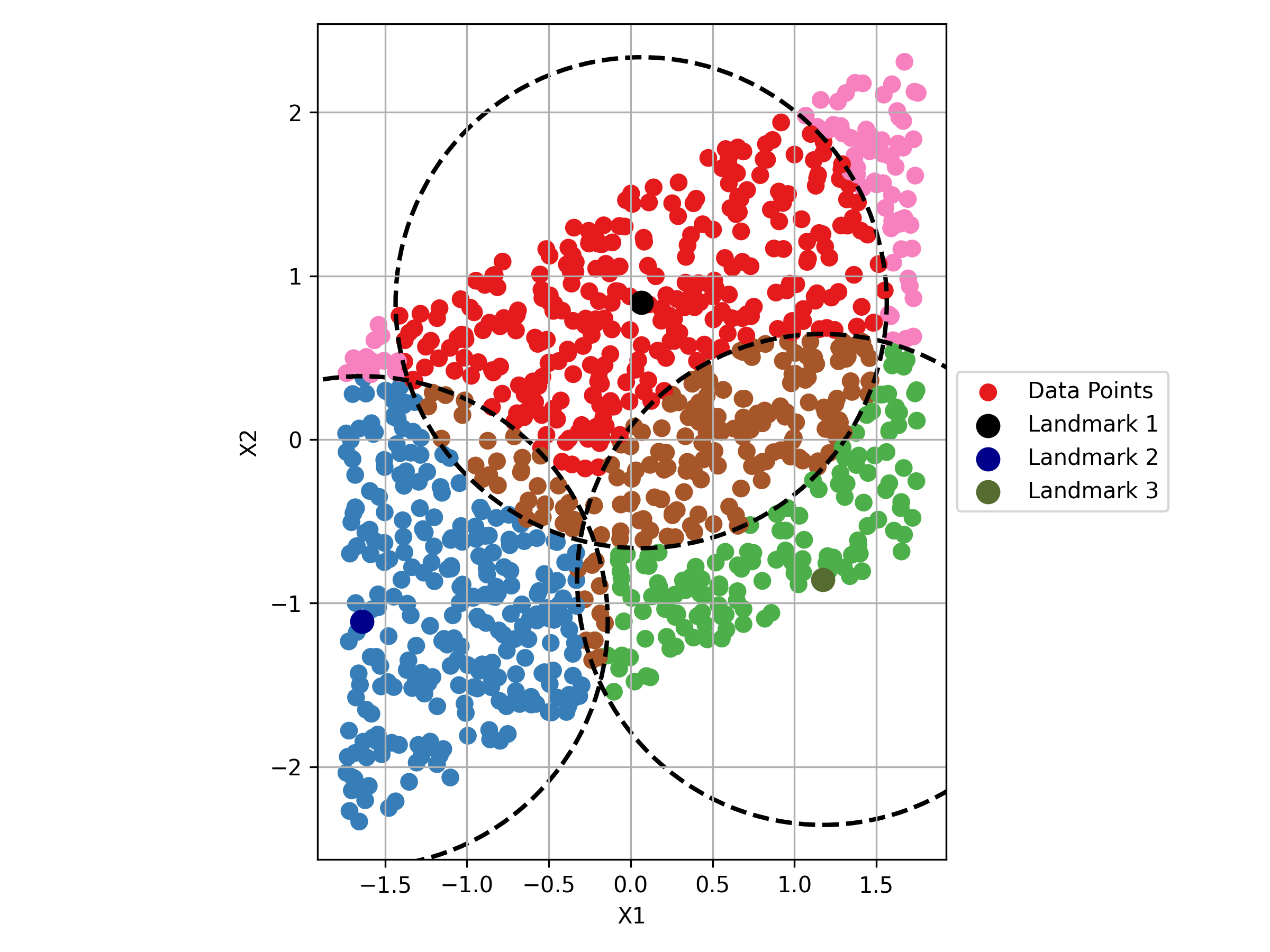}\\
			(a) Landmark 1 & (b) Landmark 2 & (c) Landmark 3  \\
			\includegraphics[width=5cm]{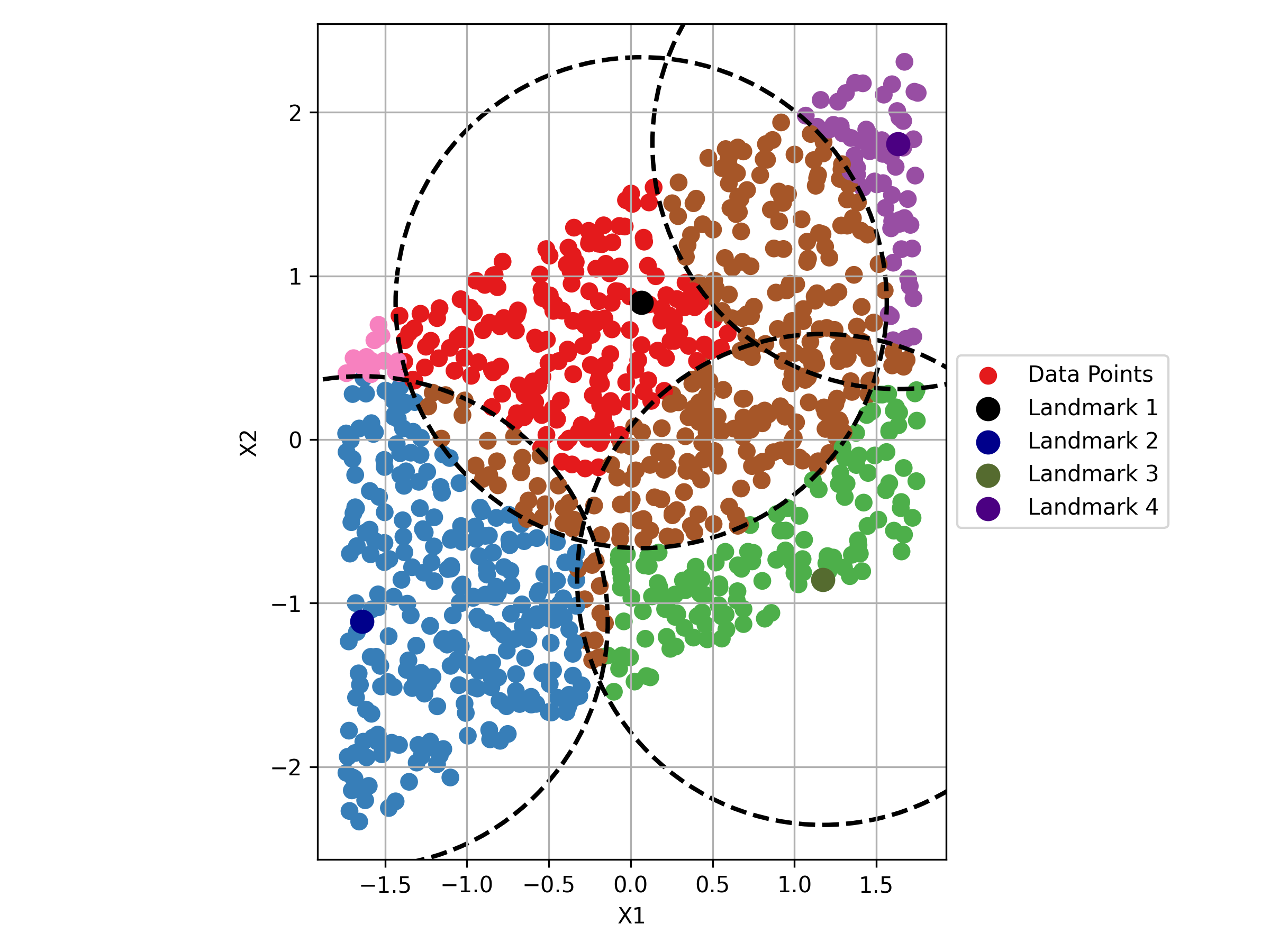}&
			\includegraphics[width=5cm]{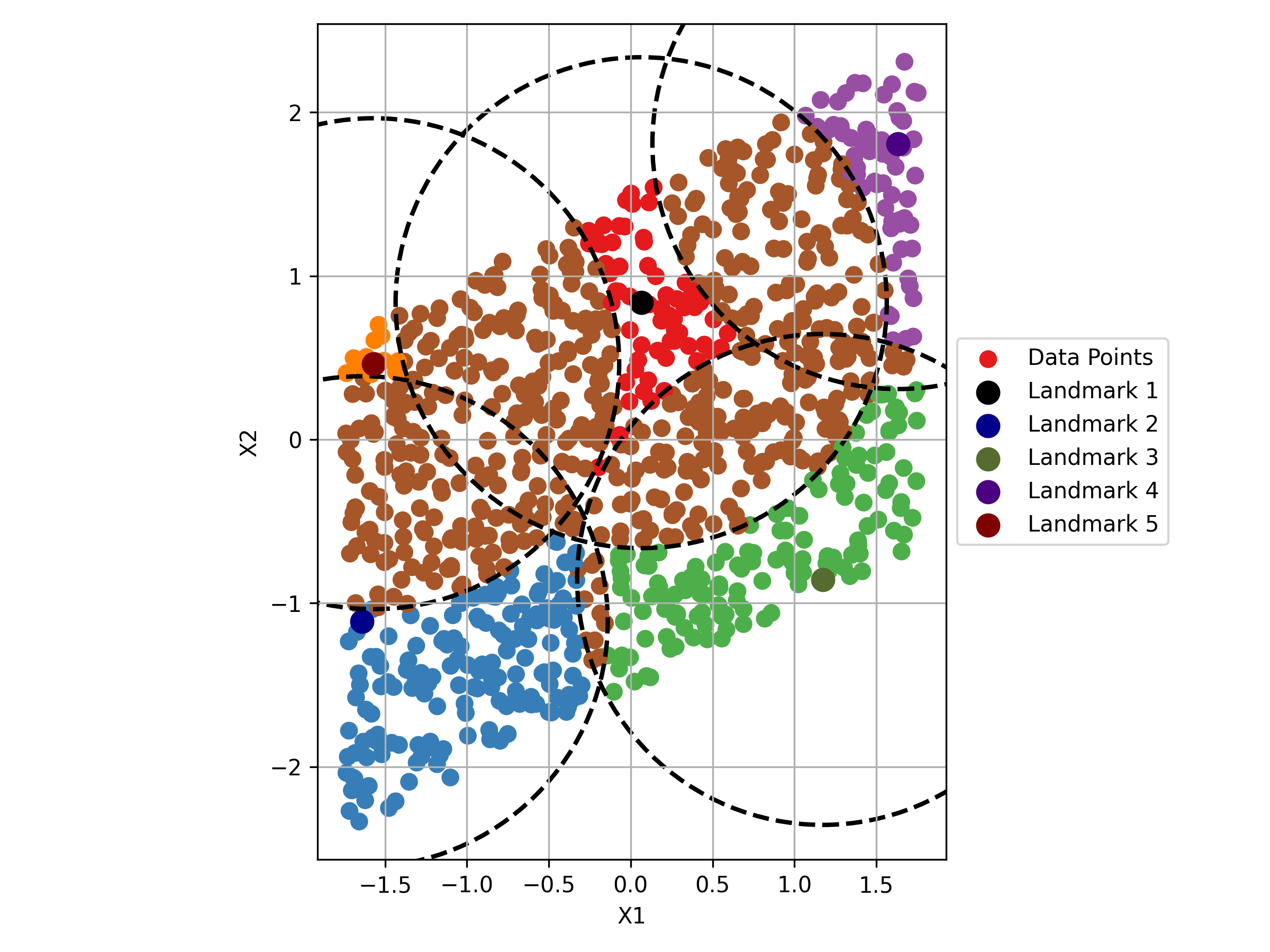}&
			\\
			(d) Landmark 4 & (e) Landmark 5 & \\
		\end{tabular}
	\end{center}
	\raggedright
	\footnotesize{Notes: Figures represent the stepwise construction of the TDABM style plot. Landmarks refer to the points that would be the landmarks of the balls in a TDABM plot. Numbering of landmarks is the order of selection. Landmarks are selected as the first index from the set of uncovered points (pink) until all points can be found within at least one circle (all other colors). Dashed circles represent the boundaries of the balls around each landmark. Dataset 2 has a correlation between $X_1$ and $X_2$ of 0.497. Dataset 2 contains 1000 points initially drawn at random from $U \left[0,1\right]$. Standardization is applied and Dataset 2 transformed to obtain the desired correlation of approximately 0.5.}
\end{figure}

For Dataset 2, comparison may also be drawn between the random selection of Figure \ref{fig:data2t} and the sequential selection shown in Figure \ref{fig:data2b}. The core difference appears with the selection of $l_3$. After the selection of $l_3$ in the sequential case there are only two pockets of uncovered points. The lower values of $X_2$ that are observed at each $X_1$ are all covered by either Ball 2 or Ball 3. Hence only 5 balls are needed to complete the cover, compared to 6 balls in the previous section. 

\begin{figure}
	\begin{center}
		\caption{Towards the TDABM Style Graph: Sequential}
		\label{fig:tbm2}
		\begin{tabular}{c c}
			\includegraphics[width=7cm]{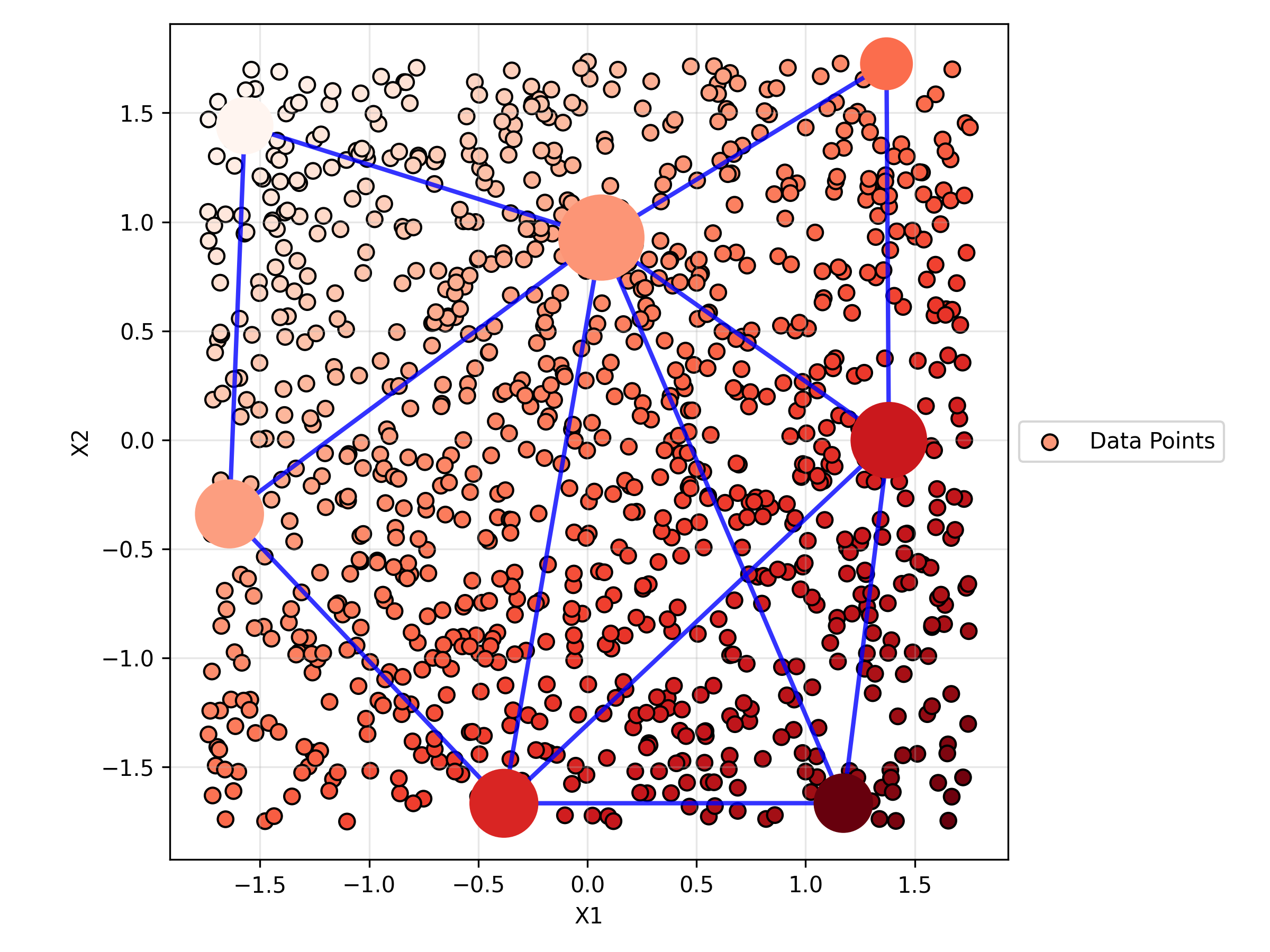}&
			\includegraphics[width=7cm]{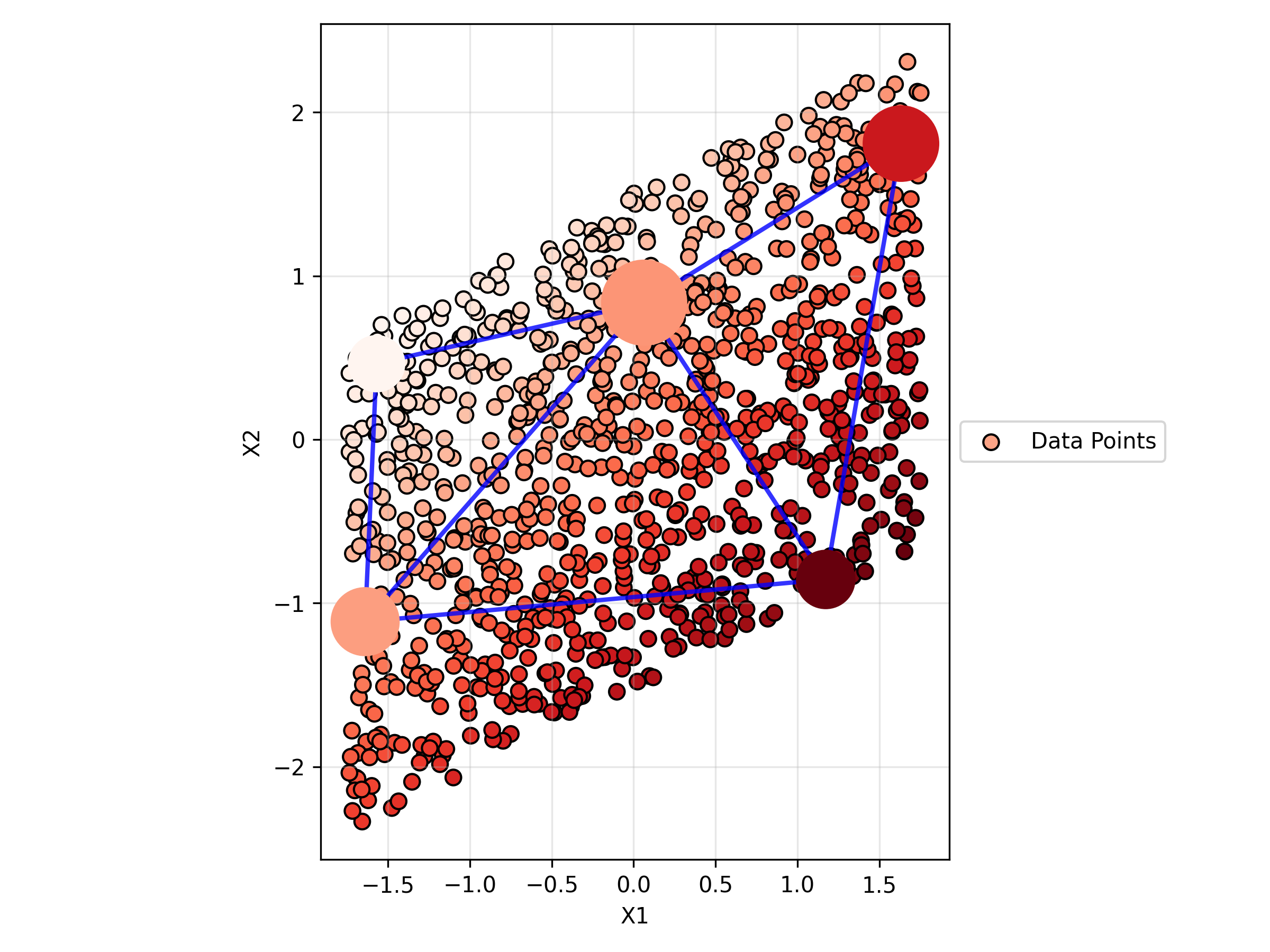}\\
			(a) Dataset 1 with Discs  & (b) Dataset 2 with Discs \\
			\includegraphics[width=7cm]{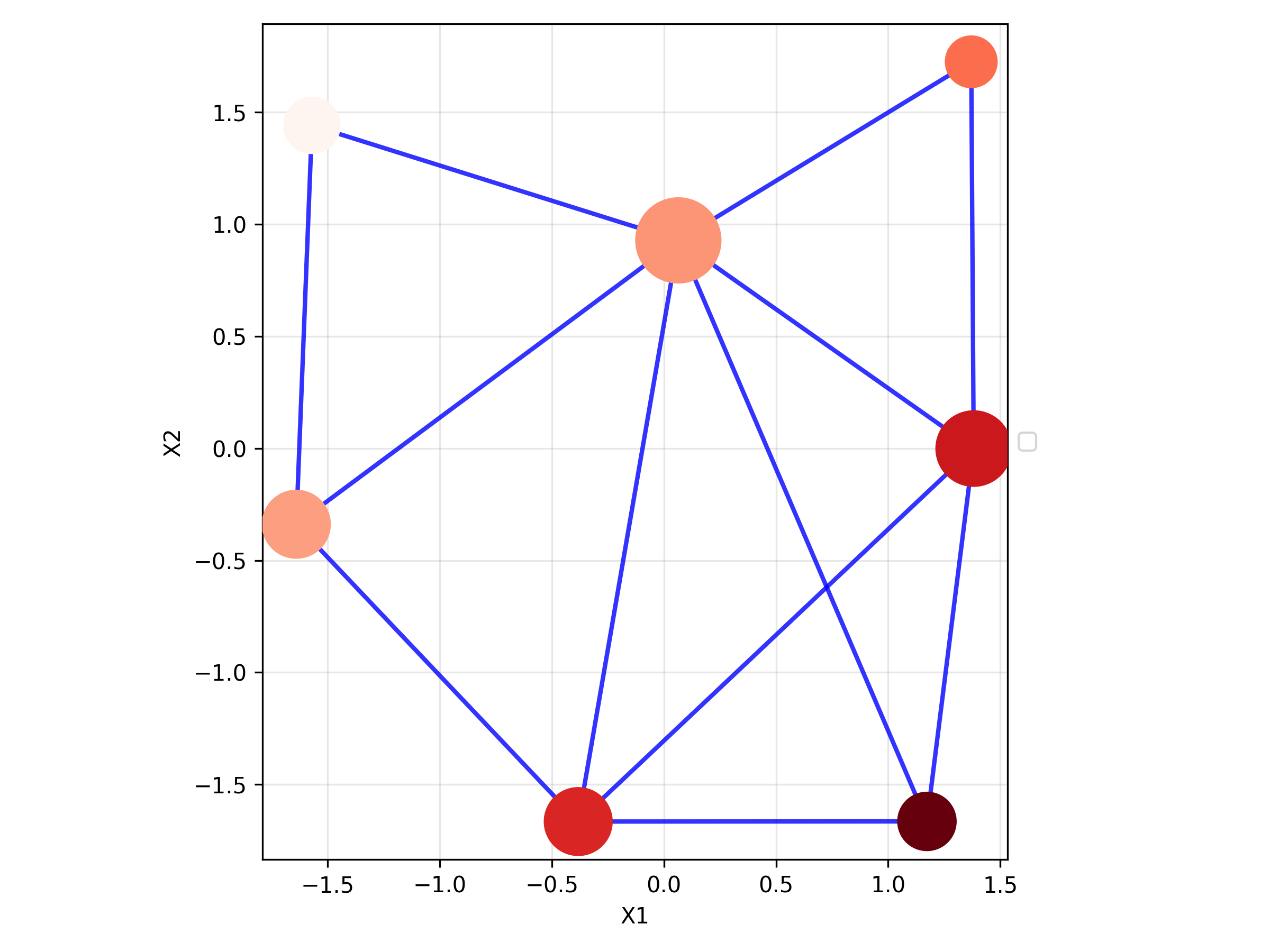}&
			\includegraphics[width=7cm]{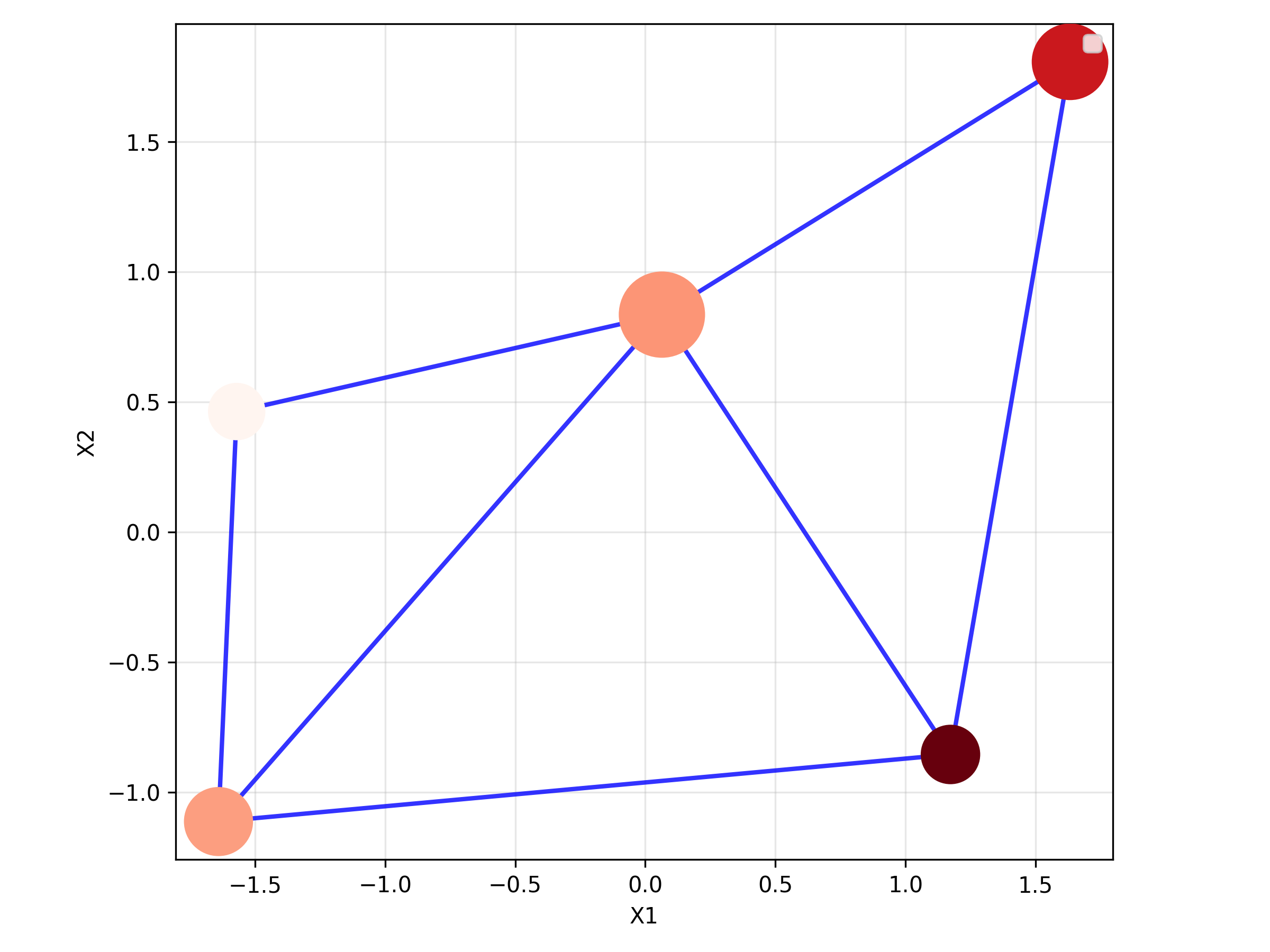}\\
			(c) Structure with Axes 1 & (d) Structure with Axes 2 \\
			\includegraphics[width=7cm]{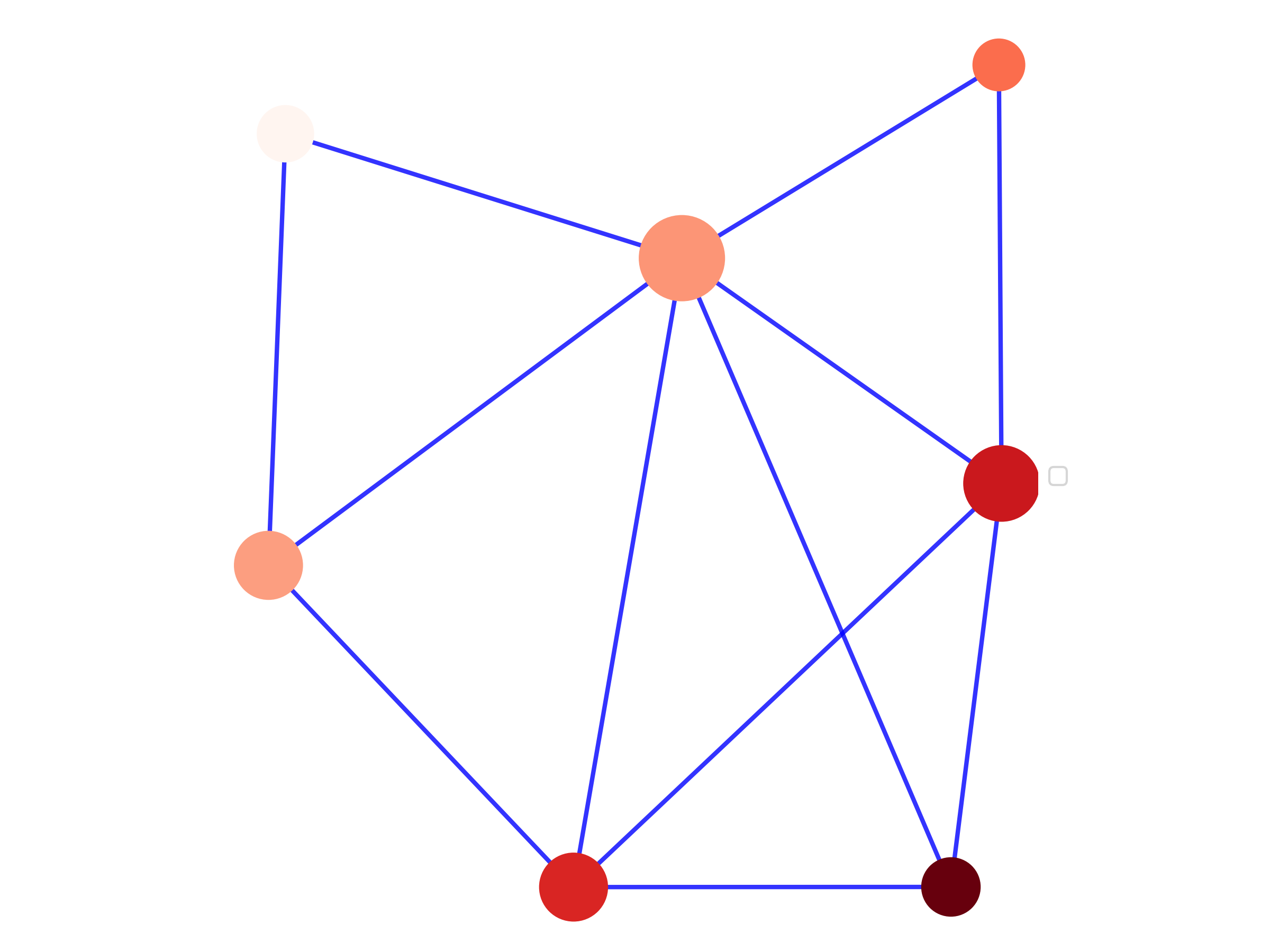}&
			\includegraphics[width=7cm]{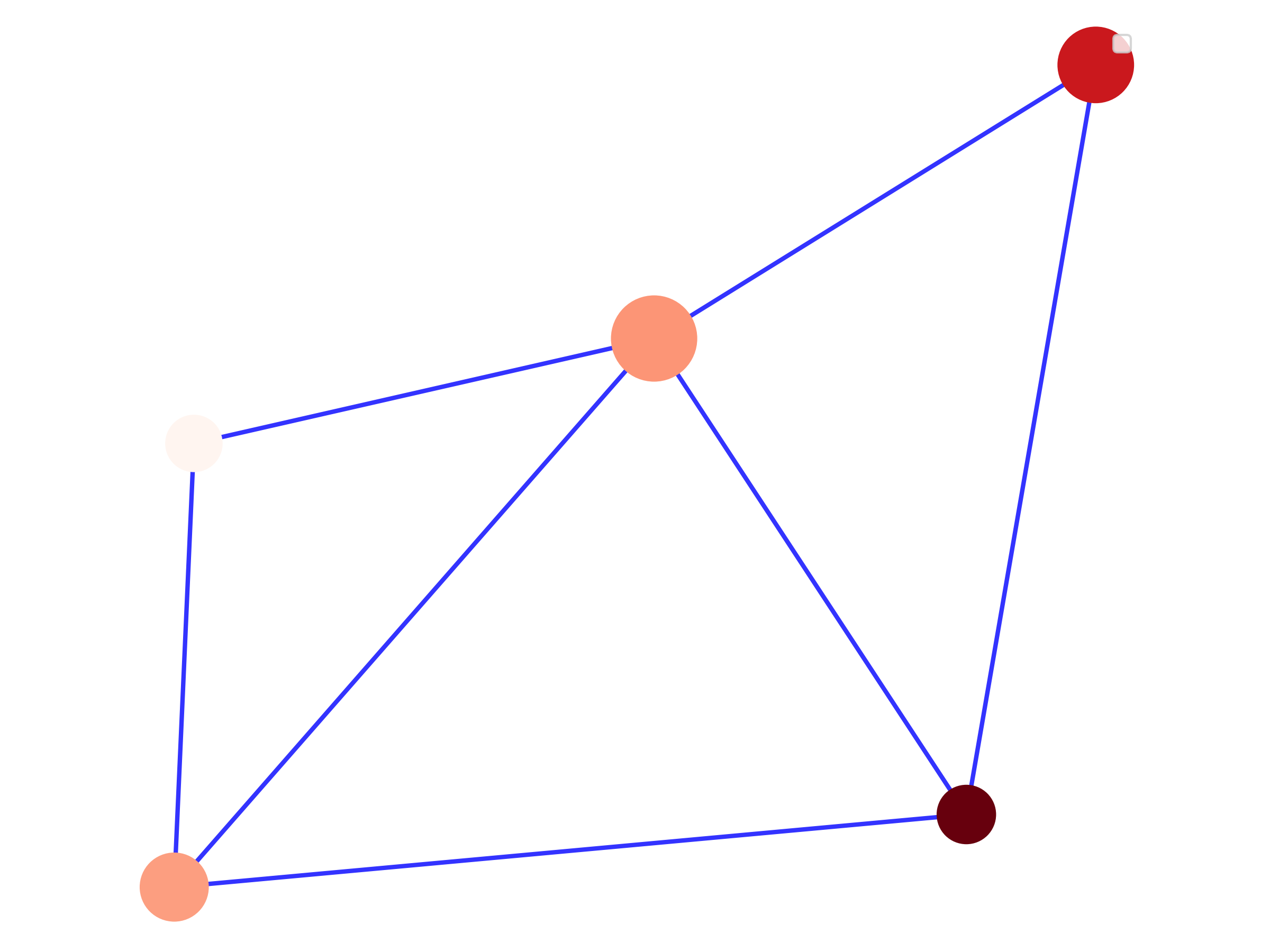}\\
			(e) TDABM Equivalent 1 & (f) TDABM Equivalent 2 \\
		\end{tabular}
	\end{center}
	\raggedright
	\footnotesize{Notes: Figures represent the transition from the full dataset to the TDABM style plot. Dataset 1 has no correlation between $X_1$ and $X_2$, whilst Dataset 2 has a correlation between $X_1$ and $X_2$ of 0.497. Both datasets contain 1000 points initially drawn at random from $U \left[0,1\right]$. Standardization is applied and Dataset 2 transformed to obtain the desired correlation of approximately 0.5.}
\end{figure}

As in the theoretical exposition, the information within the data can be removed to leave the TDABM graph. Figure \ref{fig:tbm2} shows the covers from both Dataset 1 and Dataset 2. Firstly, panels (a) and (b) plot the two datasets with each data point colored by the $Y$ value. The discs which represent each of the balls are colored according to the average value of $Y$ within the ball. The discs are sized according to the number of points within the ball. Edges are shown on the TDABM graph for the cases where the intersection between balls was non-empty. 

Panels (c) and (d) show the plots without the data points. The structure of the data can be seen in each case. The highest values of $Y$ are observed to the lower right of the data and this can be seen clearly from the plot. The graduation from the lowest values in the top left, to the highest is emphasized by the lighter diagonal then darker diagonal of balls that run from bottom left to top right. In panel (d) the graduation is less immediate, but is still clear. Once the axes are removed, the TDABM graph still retains its informativeness. However, the ability to make such quick comparison between the known properties of the data and the TDABM plot only arises because this is two-dimensional data.

The process of constructing the TDABM plots has highlighted that there are many possible plots depending on the choice of landmarks. Having multiple representations should not be seen as a major concern, inference from TDABM plots is consistent. In previously published papers it has been common to verify the consistency of inference over 1000, or 10000, random re-orderings of the data. Consistency of inference is verified in  \cite{dlotko2022topological}, \cite{rudkin2023economic}, \cite{rudkin2024return} and \cite{rudkin2024topology} amongst many others. Where a specific conclusion is hinged on the TDABM plot, such as the fact that balls may contain both the highest and lowest $Y$ values, the presence of such balls can be tested in each of the iterations. \cite{rudkin2023regional} shows that high, and low, resilience regions appear in balls which have very different historic growth trajectories. Users are encouraged to try different radii, and to try a few re-orderings of the dataset, to ensure that the results taken forward are broadly consistent.

\section{PyBallMapper}
\label{sec:pybm}

This section focuses on the implementation of TDABM in Python. I consider the command, the plotting of TDABM graphs and discuss some options for customisation. This section also discusses the use of the ball membership in analysis.

\subsection{PyBallMapper Command}

TDABM is implemented in Python using the pyBallMapper library. The core command required is the \texttt{BallMapper()} command. The inputs to the command are the $X$ variables which will be used to construct the point cloud, the $\epsilon$ ball radius, and the $Y$ variable which will be used for coloration. The preparation of the data from Dataset 1 is included in Box \ref{box:dprep}. Dataset 2 is prepared in a similar fashion.

\begin{mybox}[label=box:dprep]{Preparing Data for \texttt{BallMapper()}}
	Extract the relevant columns from the datasets. 
	\begin{lstlisting}[language=Python]
		xv1 = df1[['X1','X2']]
		y1 = df1[['Y']]
	\end{lstlisting}
\end{mybox}
The \texttt{BallMapper()} command has three input arguments. The $Y$ variable is entered as a \texttt{coloring\_df} to refer to the fact that $Y$ provides the coloration of the TDABM graph. It is possible to produce a TDABM graph without coloration if the target is simply to see the structure of the data. In most cases coloration is beneficial. Box \ref{box:pybm1} provides the code for generating a \texttt{BallMapper} object.

\begin{mybox}[label=box:pybm1]{Preparing Data for \texttt{BallMapper()}}
	Extract the relevant columns from the datasets. 
	\begin{lstlisting}[language=Python]
		bm1=pbm.BallMapper(X=xv1, eps=1.5, coloring_df=y1)
	\end{lstlisting}
\end{mybox}


The \texttt{pyBallMapper} package produces a very limited set of elements within the \texttt{BallMapper} object. Instead there are a further set of functions which will allow the user to develop additional value. The most useful is the \texttt{points\_and\_balls()} function which creates a table of all of the points in each of the balls. The point number in the table is the index from the underlying dataset. It is then possible to merge the table with the underlying dataset to construct summaries on the balls. I return to the functionality on the \texttt{BallMapper} object after the plotting section.

\begin{table}
	\begin{center}
		\caption{Elements of a \texttt{pyBallMapper} Object}
		\label{tab:bmo1}
		\begin{tabular}{|l| p{10cm}|}
			\hline
			Element & Description \\
			\hline
			\texttt{add\_coloring} & Function to allow the user to change the coloring of the TDABM plot. The inputs to the \texttt{add\_coloring()} should have a value for every data point \\ 
			\hline
			\texttt{color\_by\_variable}& Allows the user to specify one of the $X$ variables to then update the coloration of the TDABM graph\\
			\hline
			\texttt{draw\_networkx}&Used for the plotting of the TDABM graph.\\
			\hline
			\texttt{eps} & The ball radius, $\epsilon$, used in the construction of the plot\\
			\hline
			\texttt{filter\_by}& Function allowing the subsetting of the TDABM graph. For example filtering to only include the balls where $Y>0$.\\
			\hline
			\texttt{points\_and\_balls}& Creates a table with a row for each ball-point pair. The point value is the index in the $X$ data frame and allows merging back to the underlying dataset.\\
			\hline
			\texttt{points\_covered\_by\_landmarks}&A list of points in each ball. This object is the basis for the \texttt{points\_and\_balls()} function \\
			\hline
		\end{tabular}
	\end{center}
	\raggedright
	\footnotesize{Notes: List of elements which are callable after constructing a TDABM object using the function \texttt{BallMapper()} in Python. All elements are functions, with the exception of \texttt{eps} and \texttt{points\_covered\_by\_landmarks}.}
\end{table}

\subsection{Plotting a TDABM Graph}

The plotting of TDABM graphs in Python makes use of the \texttt{networkx} library. The TDABM graph is simply a set of nodes to represent the balls and edges between the nodes. The plotting function confirms the coloration variable used in the generation of the original object. There is also an option to turn on, or off, the colorbar at the side of the plot. The colorbar is used to allow the reader to understand the values being represented by the colors of the disks in the plot. Box \ref{box:pybm3} provides the code.

\begin{mybox}[label=box:pybm3]{Plotting the TDABM graph}
	Plotting is based upon a \texttt{BallMapper} object which has already been created
	\begin{lstlisting}[language=Python]
		bm1.draw_networkx(coloring_variable='Y', colorbar=True)
	\end{lstlisting}
\end{mybox}

\begin{figure}
	\begin{center}
		\caption{TDABM Plots and Comparisons}
		\label{fig:tdabm}
		\begin{tabular}{c c}
			\includegraphics[width=7cm]{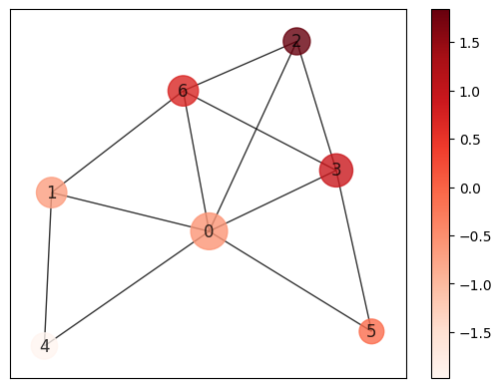}&
			\includegraphics[width=7cm]{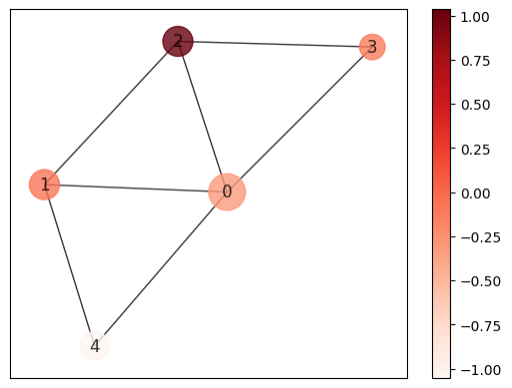}\\
			(a) TDABM Graph 1 & (b) TDABM Graph 2 \\
			\includegraphics[width=7cm]{bmsd1m03.png}&
			\includegraphics[width=7cm]{bmsd2m03.png}\\
			(e) TDABM Equivalent 1 & (f) TDABM Equivalent 2 \\
		\end{tabular}
	\end{center}
	\raggedright
	\footnotesize{Notes: TDABM Graphs and corresponding step-by-step examples from the previous section of this guide. Dataset 1 has no correlation between $X_1$ and $X_2$, whilst Dataset 2 has a correlation between $X_1$ and $X_2$ of 0.497. Both datasets contain 1000 points initially drawn at random from $U \left[0,1\right]$. Standardization is applied and Dataset 2 transformed to obtain the desired correlation of approximately 0.5. }
\end{figure}

The plots generated for Dataset 1 and 2 are compared with the step-by-step examples generated in the previous section. The aim here is to show that the intuition from the walk-through example applies when the algorithm is allowed to generate the TDABM graph directly. Figure \ref{fig:tdabm} contains the relevant four plots. Coloration in all cases is by the average $Y$ value within the ball. 

Figure \ref{fig:tdabm} shows that there are very strong similarities between the graphs produced by the \texttt{bm1.draw\_networkx()} and the graphs that were generated when going step-by-step through the implementation. Panels (a) and (c) related to Dataset 1 and clearly indicate the structure identically. The two plots are almost symmetric, with Ball 4, to the lower left of the code generated graph being to the bottom left and then being to the upper left in the step-by-step case. Likewise, Ball 5 has a mid-range coloration and can be seen to the lower right in panel (a), but upper right in panel (c). A similar reflection is seen for Dataset 2 in panels (b) and (d). There are small variations in the coloration and sizing of the balls. The difference in color arises from the fact that the  \texttt{bm1.draw\_networkx()} adds a small buffer at the top and bottom of the coloration range. The difference in size arises because the smallest and largest disc sizes are normalized to a slightly different range by the respective plotting commands.

\begin{figure}
	\begin{center}
		\caption{TDABM Plots Recolored}
		\label{fig:tdabm2}
		\begin{tabular}{c c}
			\includegraphics[width=7cm]{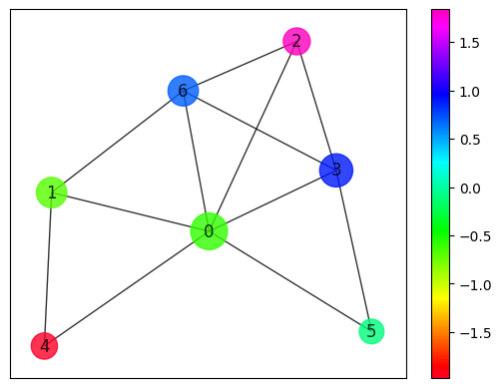}&
			\includegraphics[width=7cm]{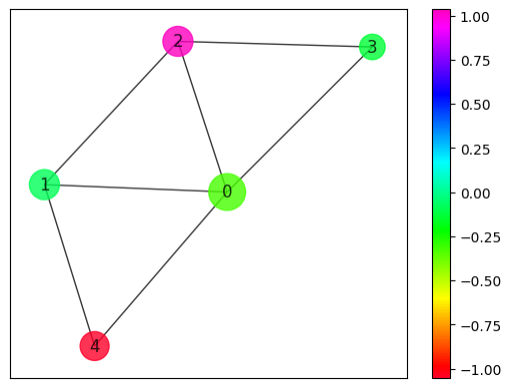}\\
			(a) TDABM Graph 1 & (b) TDABM Graph 2 \\
		\end{tabular}
	\end{center}
	\raggedright
	\footnotesize{Notes: TDABM Graphs and corresponding step-by-step examples from the previous section of this guide. Coloration is updated to use the \texttt{gist\_rainbow} colormap. Dataset 1 has no correlation between $X_1$ and $X_2$, whilst Dataset 2 has a correlation between $X_1$ and $X_2$ of 0.497. Both datasets contain 1000 points initially drawn at random from $U \left[0,1\right]$. Standardization is applied and Dataset 2 transformed to obtain the desired correlation of approximately 0.5. }
\end{figure}

The default setting within python is to use the red colormap. This guide has used the red colormap throughout as a result. However, it is straightforward to use an alternative colormap for the TDABM graphs. Box \ref{box:pybm4} shows the updated command with a \texttt{color\_map} option set. In this case a rainbow colormap is added from the colormaps library. Box \ref{box:pybm4} also includes the code to generate the colormap. In this case the colormap used is \texttt{gist\_rainbow}. The user is free to develop their own colormaps provided they satisfy the usual features of a Python colormap. Development of new colormaps is beyond the scope of this guide. The resulting TDABM graphs for Datasets 1 and 2 are plotted in Figure \ref{fig:tdabm2}.

\begin{mybox}[label=box:pybm4]{Coloring the TDABM graph}
	Continuing with the \texttt{bm1} object which has been created by the \texttt{BallMapper()} function. First the colormap is obtained
	\begin{lstlisting}[language=Python]
		hsvp = cm.get_cmap("gist_rainbow")
	\end{lstlisting}
	The graph can now be plotted
	\begin{lstlisting}[language=Python]
		bm1.draw_networkx(coloring_variable='Y', color_palette=hsvp, colorbar=True)
	\end{lstlisting}
\end{mybox}

\begin{figure}
	\begin{center}
		\caption{TDABM Plots By Axis Recolored}
		\label{fig:tdabm5}
		\begin{tabular}{c c}
			\includegraphics[width=7cm]{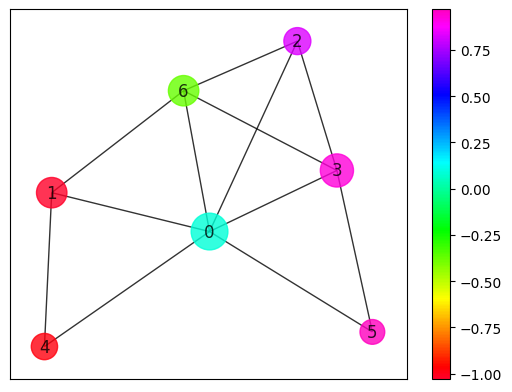}&
			\includegraphics[width=7cm]{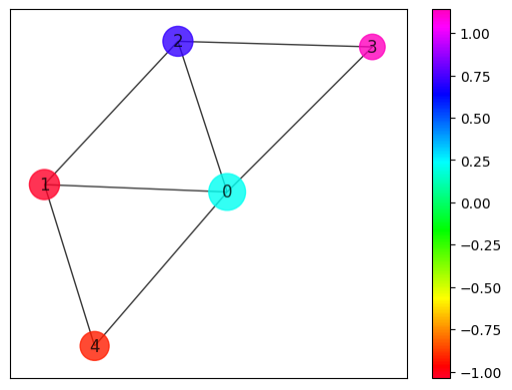}\\
			(a) Dataset 1: $X_1$ & (b) Dataset 2: $X_1$ \\
			\includegraphics[width=7cm]{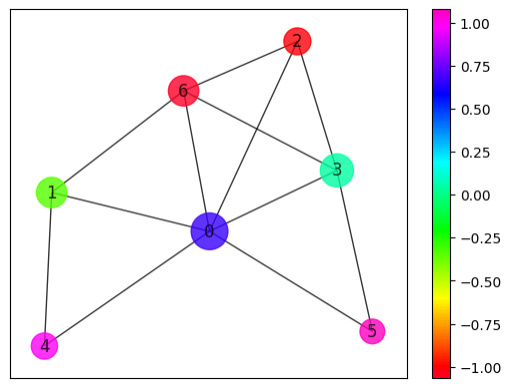}&
			\includegraphics[width=7cm]{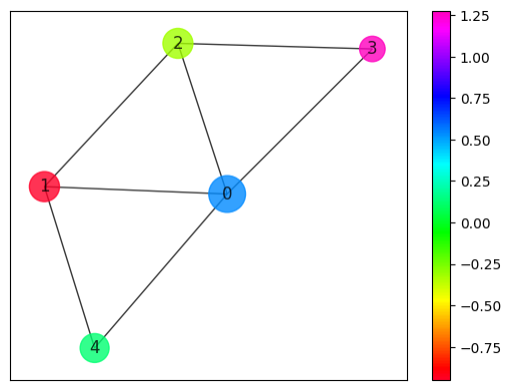}\\
			(c) Dataset 1: $X_2$ & (d) Dataset 2: $X_2$ \\
		\end{tabular}
	\end{center}
	\raggedright
	\footnotesize{Notes: TDABM Graphs colored by the two axis variables, $X_1$ and $X_2$. Coloration is updated to use the \texttt{gist\_rainbow} colormap. Dataset 1 has no correlation between $X_1$ and $X_2$, whilst Dataset 2 has a correlation between $X_1$ and $X_2$ of 0.497. Both datasets contain 1000 points initially drawn at random from $U \left[0,1\right]$. Standardization is applied and Dataset 2 transformed to obtain the desired correlation of approximately 0.5. }
\end{figure} 

Figure \ref{fig:tdabm5} provides the TDABM graphs from Figure \ref{fig:tdabm2} colored according to the two axis values. Using the $X_1$ coloration in panel (a), the fact that high $X_1$ value remain on the right, low $X_1$ values remain on the left, of the TDABM graph is clear. The $X_2$ coloration shows that higher values of $X_2$ are now found to the lower edge of the TDABM plot. This is the inversion of the $Y$ values discussed opposite Figure \ref{fig:tdabm2} in action. For Dataset 2, the plot was also inverted, the higher values of $X_2$ are again comparatively lower in the TDABM plot. The highest values of $X_1$ remain on the right of the TDABM plot. The correlated nature of the data means that the highest, and lowest, values of $X_1$ and $X_2$ appear in similar parts of the space.  

\begin{figure}
	\begin{center}
		\caption{TDABM Plots and Comparisons}
		\label{fig:tdabm3}
		\begin{tabular}{c c}
			\includegraphics[width=7cm]{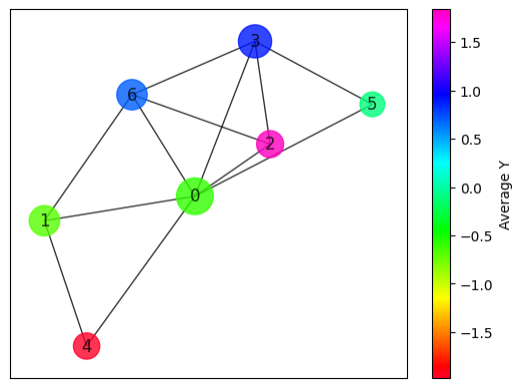}&
			\includegraphics[width=7cm]{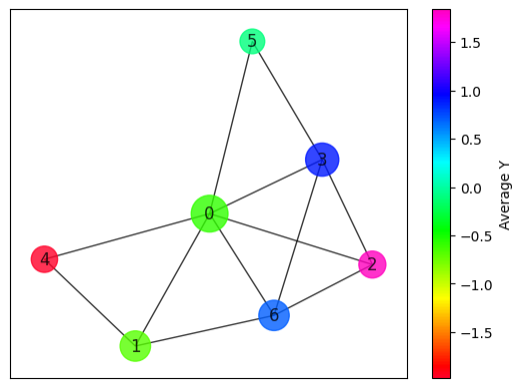}\\
			(a) TDABM Graph 1 Seed 1  & (b) TDABM Graph 1 Seed 5 \\
		\end{tabular}
	\end{center}
	\raggedright
	\footnotesize{Notes: TDABM Graphs and corresponding step-by-step examples from the previous section of this guide. Coloration is updated to use the \texttt{gist\_rainbow} colormap. Dataset 1 has no correlation between $X_1$ and $X_2$, whilst Dataset 2 has a correlation between $X_1$ and $X_2$ of 0.497. Both datasets contain 1000 points initially drawn at random from $U \left[0,1\right]$. Standardization is applied and Dataset 2 transformed to obtain the desired correlation of approximately 0.5. }
\end{figure}

The inference from Figure \ref{fig:tdabm2} is identical to that from panels (a) and (b) in Figure \ref{fig:tdabm2}. The networkx package also has features to allow the user to work on the presentation of the graphs. To handle the compression from multiple dimensions into an abstract representation in two-dimensions, a spring algorithm is used. The spring algorithm ensures that balls with greater similarity are placed nearer to each other on the page. There are two parameters of interest. The random seed, \texttt{seed}, and the extent of the spring, $k$. For the bivariate case the $k$ parameter does not make a clear difference to the plots. However, the seed can be shown to have an effect. Figure \ref{fig:tdabm3} shows two different seed values for Dataset 1.

From Figure \ref{fig:tdabm3} it can be readily seen that there is no impact created by the choice of seed. The core elements are still the set of connected balls amongst the higher values of $Y$. The full connectivity between balls 0, 2, 3 and 6 contrasts with the lack of connection between balls 4 and 6 (for example). Irrespective of where these balls appear, the edges are the same. It is important to remember that the visualization of a TDABM plot is abstract. Any seed can be used to produce the plot that best allows the user to understand the data. Before committing to a particular seed for a paper, users are encouraged to look from different seeds. 

In order to customize the plots of TDABM graphs in Python, the user can adjust several features. Table \ref{tab:plotinput} provides a summary of some of the features which can be adapted. In this guide focus has been on adjusting the coloration, the spring parameter, the seed and the label on the colorbar. Where it is intended to do comparisons across multiple plots, the setting of an identical range for the coloration is a helpful option. By setting the range the interpretation of each color is the same in each plot. 

\begin{table}
	\begin{center}
		\caption{Options for TDABM Plotting with \texttt{draw.networkx()}}
		\label{tab:plotinput}
		\begin{tabular}{|l |l |p{8cm}|}
		 \hline
		 Setting & Command & Description \\
		 \hline
		 BallMapper Object & - & The first input is always a \texttt{BallMapper} object created using the \texttt{BallMapper()} command. \\
		 \hline
		 Colormap & \texttt{cmap} & Allows the specification of the colormap used in the plotting. In the examples here predefined examples have been used.\\
		 \hline
		 Colorbar & \texttt{colorbar} & A True/False indicator to state whether the colorbar will be displayed. Default is True.\\
		 \hline
		 Colorbar Label & \texttt{colorbar\_label} & Allows the user to specify the labelling for the colorbar. Useful where the variable name and the actual name are different. \\
		 \hline
		 Seed for plotting & \texttt{seed} &  Because Python is trying to place balls from a higher dimension onto the page, there is natural scope for variation in the placement. The seed parameter allows the user to control the placement and ensures reproducibility. Set within the \texttt{pos=nx.spring\_layout(bm1.Graph, k=1, seed=5))} argument \\
		 \hline
		 Spring parameter & \texttt{k} & Sets the extent to which balls which have overlap are attracted to each other. Lower values cause connected groups of balls to stick closer together. Lower values may cause the connected balls to extend wider than the disconnected balls. Set within the \texttt{pos=nx.spring\_layout(bm1.Graph, k=1, seed=5))} argument\\
		 \hline
		 Minimum Color & \texttt{vmin} & Sets the lowest value on the colorbar. Default is the lowest valuation of the coloration variable minus a small allowance \\
		 \hline
		 Maximum Color & \texttt{vmax} & Sets the highest value on the colorbar. Default is the highest valuation of the coloration variable plus a small allowance \\
		 \hline
		\end{tabular}
	\end{center}
	\raggedright
	\footnotesize{Notes: Elements which can be passed to the \texttt{draw.networkx()} command in the plotting of the TDABM graph. The second column provides the argument in the function. For example, to change the colormap to the \texttt{viridis} defined map, the argument would be \texttt{cmap = 'viridis'} . The seed for plotting and spring parameter are set within a sub-function. To change the seed for plotting and the spring parameter adjust within the \texttt{pos=nx.spring\_layout(bm1.Graph, k=1, seed=5))} argument.}
\end{table}

Advanced users may wish to develop additional elements of the plot. It is possible to change transparency, edge thickness and text settings. It is also possible to create user-defined colormaps to convey specific messages more clearly. \cite{rudkin2023economic} create a bespoke colormap for their work on voting patterns which splits at 50\%. The 50\% split represents the margin needed for the vote to Leave the European Union to exceed the vote to Remain. \cite{rudkin2023economic} use a blue graded colormap for the Remain vote and a contrasting orange colormap for ball colors above 50\%. The ability to adjust parameters within the plotting of TDABM graphs creates further scope for advancing the value of TDABM. 

\subsection{Ball Membership}

An advantage of the TDABM algorithm is that it is always possible to refer back to individual datapoints within the balls. Because the algorithm constructs a cover, each ball has a list of members available for interrogation. In order to obtain the list of ball members from the Python implementation, the \texttt{points\_and\_balls()} function is applied. The result is a table of dataset index numbers that are contained within each ball. Because there are points in the intersection of pairs of balls, some data points appear more than once in the newly created table. The point number is the index from the X data that was supplied to the TDABM algorithm. Having a version of the input data set with the index as a variable is helpful. Box \ref{box:pybm4} contains the required code.

\begin{mybox}[label=box:pybm4]{Ball Membership}
	First the \texttt{points\_and\_balls()} function is applied to the \texttt{BallMapper} object
	\begin{lstlisting}[language=Python]
		bm1a = bm1.points_and_balls()
	\end{lstlisting}
	To merge a variable called \texttt{point} is needed within the dataframe. It is vital that the index is created at the same time as the $X$ variable is extracted. If the data is re-ordered at any stage then the indexes will no longer match.
	\begin{lstlisting}[language=Python]
		df1a = pd.merge(df1,bm1a,on="point")
	\end{lstlisting}
\end{mybox}

The dataset that is created, \texttt{df1a}, has more rows than \texttt{df1} because of the points within the intersection. The resulting expanded dataframe can be queried to understand exactly which points are within which balls. For example, \cite{rudkin2023economic} use the ability to identify points within balls to understand which parliamentary constituencies are in each of their constituency characteristic balls. In this case, the data is artificial and so the points have no particular meaning. To see details for a single ball a subsetting can be applied. Here summary statistics are constructed for each ball. The resulting summary statistics are in Table \ref{tab:summary}.

\begin{sidewaystable}
	\begin{center}
		\caption{Summary of Balls}
		\label{tab:summary}
		\begin{tabular}{l l c c c c c c c c c c c c c }
			\hline
			Dataset & Ball & \multicolumn{4}{l}{Variable: $X_1$} & \multicolumn{4}{l}{Variable: $X_2$} & \multicolumn{4}{l}{Variable: $Y$} & Obs \\
			&& Mean & sd & Min & Max & Mean & sd & Min & Max & Mean & sd & Min & Max & \\
			\hline
											
			1 &0&	0.085&	0.772&	-1.429&	1.542&	0.672&	0.617&	-0.531&	1.734&	-0.587&	0.995&	-2.768&	1.255 & 485\\
			&1&	-1.028&	0.438&	-1.738&	-0.180&	-0.340&	0.746&	-1.748&	1.113&	-0.688&	0.880&	-2.784&	0.811 & 307\\
			&2&	0.826&	0.554&	-0.314&	1.748&	-1.012&	0.454&	-1.748&	-0.182&	1.838&	0.614&	0.759&	3.400 & 225\\
			&3&	0.931&	0.505&	-0.082&	1.753&	0.044&	0.750&	-1.451&	1.487&	0.887&	0.893&	-0.687&	3.086 & 380\\
			&4&	-0.985&	0.487&	-1.738&	-0.112&	0.991&	0.451&	0.002&	1.707&	-1.976&	0.587&	-3.244&	-0.915 & 209\\
			&5&	0.971&	0.492&	-0.069&	1.753&	1.076&	0.407&	0.294&	1.734&	-0.105&	0.697&	-1.734&	1.365 & 176 \\
			&6&	-0.373&	0.723&	-1.736&	1.059&	-1.065&	0.414&	-1.749&	-0.195&	0.693&	0.857&	-0.814	&2.639& 306\\
			2& 0 &	0.215 &	0.727&	-1.412&	1.552&	0.540&	0.590&	-0.614&	1.937&	-0.325&	0.834&	-2.168&	1.309 & 503\\
			&1&	-1.038&	0.449&	-1.738&	-0.157&	-0.950&	0.645&	-2.334&	0.375&	-0.088&	0.803&	-2.007&	1.273 & 309\\
			&2&	0.718&	0.549&	-0.323&	1.748&	-0.317&	0.548&	-1.542&	0.597&	1.035&	0.558&	-0.058&	2.338 & 307\\
			&3&	1.140&	0.401&	0.254&	1.753&	1.272&	0.494&	0.355&	2.308&	-0.132&	0.619&	-1.278&	1.223 & 199\\
			&4&	-0.926&	0.474&	-1.738&	-0.104&	0.126&	0.548&	-1.026&	1.294&	-1.053&	0.589&	-2.240&	0.029 & 280\\
			
			\hline
		\end{tabular}		
	\end{center}
	\raggedright
	\footnotesize{Notes: Summary statistics for the balls in the TDABM plots in Figure \ref{fig:tdabm}. The axis variables are $X_1$ and $X_2$. The coloration is $Y$. Across the whole dataset $Y = X_1 - X_2$. The number of points in each ball is given in the final column.}
\end{sidewaystable}

Table \ref{tab:summary} shows how each ball sits in a different part of the $X_1, X_2$ space. The maximum difference between the highest, and lowest, values of $X_1$ within the same ball is 3. A difference of 3 occurs when there are two points on the edges of the ball with exactly the same value for $X_2$. Some balls are close to this, for example Ball 0 in Dataset 1. A radius of 1.5 in standardized data does ensure that the balls have large coverage. The largest ball in Dataset 2 covers more than half of the data points, whilst the largest ball in Dataset 1 has just under half at 485. In the context of an artificial dataset, the summary statistics have limited interpretation. However, in application to a real-world dataset these summary statistics assist in understanding. 

\section{Summary}
\label{sec:summary}

Topological Data Analysis Ball Mapper (TDABM) \citep{dlotko2019ball} is a means for visualizing multi-dimensional data in an abstract two-dimensional form. This guide has provided an introduction to the implementation of the TDABM algorithm in Python using the \texttt{PyBallMapper} library. As a start point, data is seen as a point cloud. TDABM creates a cover of the point cloud by sequentially selecting an uncovered datapoint and surrounding that landmark point with a ball of radius $\epsilon$. The algorithm continues selecting landmarks from the uncovered set until all datapoints are members of at least one ball. Two simulations of the algorithm have shown that the order of landmark selection is important to the specific appearance of the graph. However, because the location of each data point within the point cloud is fixed there is an appealing consistency to the inference on all realizations of the random landmark selection.

The power of TDABM as a model free means of visualization has seen increased applications, including the recent work of \cite{otway2024shape, rudkin2023economic, rudkin2024return, tubadji2025cultural}, and many others. By appreciating the structure within the dataset it is possible to say more about the data which is then informing any empirical study. As demonstrated by \cite{dlotko2021financial}, \cite{rudkin2024return} and others, the TDABM plot also offers an opportunity to comment on the quality of model fit. The importance of visualization at the start, and end, of the modelling process is emphasized by \cite{anscombe1973graphs}, \cite{matejka2017same} and many others. TDABM represents a valuable contribution to the visualization task.

There are alternative means to visualize multi-dimensional data. There is an ongoing research agenda to provide direct comparisons with approaches such as UMAP, T-SNE and the original mapper of \cite{singh2007topological}. Relative to the alternative approaches, the TDABM algorithm has the advantage of not requiring any data reduction, and of only having a single choice parameter. TDABM also has the ability to group observations within a ball. However, the aim of TDABM is not to create a clustering. The balls are simply all points within a given radius of the landmark point. Clustering algorithms, such as k-means, target their choice parameters on the within group to between group variation ratio. Comparisons between TDABM and clustering have been a feature of papers applying TDABM such as \cite{rudkin2023economic}, \cite{otway2024shape} and others. Users are encouraged to evaluate the best approach for their own settings. However, to date the TDABM algorithm has proven to be informative on all datasets studied.

The PyBallMapper implementation is actively maintained by the Dioscuri Centre in Topological Data Analysis, and the GitHub associated with the project is regularly updated\footnote{The GitHub site is: \href{https://github.com/dioscuri-tda/pyBallMapper/blob/main/README.md}{https://github.com/dioscuri-tda/pyBallMapper/blob/main/README.md}.}. Interested users may find further resources on TDABM on the Disocuri Centre GitHub. This guide has been written to allow an accessible route into the power of TDABM. Whether for model assessment, or simply understanding the structure of data, TDABM offers a valuable addition to the data analysis toolkit.

\bibliography{tdabmpres}
\bibliographystyle{apalike}

\end{document}